\documentclass[twocolumn,numberedappendix]{emulateapj}
\usepackage{apjfonts}
\usepackage{natbib}

\usepackage{graphicx}

\bibpunct[,]{(}{)}{;}{a}{}{,}

\def\totd{{\mathrm{d}}}

\begin{document}

\title{The X-Ray Polarization Signature of Quiescent Magnetars: Effect of Magnetospheric Scattering and
       Vacuum Polarization}
\author{Rodrigo Fern\'andez\altaffilmark{1,2} and Shane W. Davis\altaffilmark{1,3}}
\altaffiltext{1}{Institute for Advanced Study. Einstein Drive, Princeton, NJ 08540, USA.}
\altaffiltext{2}{Einstein Fellow}
\altaffiltext{3}{Canadian Institute for Theoretical Astrophysics. Toronto, ON M5S3H4, Canada.}

\begin{abstract}
In the magnetar model, the quiescent non-thermal soft X-ray emission from Anomalous X-ray Pulsars
and Soft-Gamma Repeaters is thought to arise from resonant comptonization of thermal photons by charges
moving in a twisted magnetosphere. 
Robust inference of physical quantities from observations is difficult, because the process 
depends strongly on geometry and current understanding of the magnetosphere is not very deep. 
The polarization of soft X-ray photons is an
independent source of information, and its magnetospheric imprint 
remains only partially explored.
In this paper we calculate how resonant cyclotron scattering would modify
the observed polarization signal relative to the surface emission, using a multidimensional
Monte Carlo radiative transfer code that accounts for the gradual coupling of polarization eigenmodes
as photons leave the magnetosphere. We employ a globally-twisted, self-similar, force-free 
magnetosphere with a power-law momentum distribution, assume a blackbody 
spectrum for the seed photons, account for general relativistic light deflection close to the star, 
and assume that vacuum polarization dominates the dielectric properties of the magnetosphere.
The latter is a good approximation if the pair multiplicity is not much larger than unity.
Phase-averaged polarimetry is able to provide a clear signature of the magnetospheric
reprocessing of thermal photons and to constrain mechanisms generating the thermal emission.
Phase-resolved polarimetry, in addition, can characterize the spatial extent and magnitude of
the magnetospheric twist angle at $\sim 100$ stellar radii, and discern between uni- or bidirectional 
particle energy distributions, almost independently of every other parameter in the system. We discuss 
prospects for detectability with GEMS.
\end{abstract}

\keywords{radiative transfer --- magnetic fields --- plasmas --- stars: neutron --- X-rays: stars --- techniques: polarimetric}

\maketitle

\section{Introduction}

The rich phenomenology associated with Soft Gamma Repeaters (SGRs; e.g. \citealt{mereghetti2007}) 
and Anomalous X-ray Pulsars (AXPs; e.g. \citealt{kaspi2007}) has so far been best accounted 
for by the \emph{magnetar} model, which hypothesizes the decay of a strong magnetic field in a neutron star as the dominant 
energy source \citep{duncan1992,thompson1996}.

In the soft X-ray band ($1$--$10$~keV), the quiescent spectra of 
magnetar candidates are, to first approximation, very similar:
a thermal component with temperature $kT\sim 0.4$~keV, 
and a non-thermal tail of varying hardness at higher energies
(e.g., \citealt{woods2006}). On an individual basis, though,
quiescent AXPs and SGRs can show variability in their spectra
and pulse profiles over timescales ranging from days to
years, in many cases around outburst periods or
glitches (e.g., \citealt{kaspi2007,woods2007}). 

In the magnetar model, the non-thermal component of the spectrum arises from the decay of
the toroidal part of the magnetospheric field (the ``twist"), 
thought to be acquired during outbursts \citep{thompson2002}. The twist induces a current
along the field lines, increasing the density of charges by orders of magnitude above
the \citet{goldreich1969} value, resulting in an optical depth to resonant
cyclotron scattering of order unity. Comptonization of thermal photons from the stellar surface
then generates a non-thermal tail in the spectrum, and pulsed emission from               
the combination of stellar rotation and the geometric dependence of the process \citep{thompson2002}.

Numerical calculations of the resulting emission
have been carried out with both one-dimensional
semianalytic methods \citep{lyutikov2006} and multidimensional Monte Carlo 
algorithms \citep{fernandez2007,nobili2008a}. Output from the one-dimensional 
computation has been used to model phase averaged spectra of AXPs and SGRs \citep{guver2007,guver2008, 
rea2007b,rea2008}, 
resulting in good fits while improving upon the ad-hoc blackbody plus power-law prescription. 
Even better fits are obtained when using the results of the \citet{nobili2008a} calculation,
as no extra power-law is required to fit SGR spectra \citep{zane2009}.

Still, phase averaged observables convey only part of the information, raising the question of geometric degeneracies.
Recently, \citet{albano2010} have performed simultaneous light curve and spectral fitting
of the transient AXPs XTE J1810-197 and CXOU J164710.2-455216, making use of the \citet{nobili2008a} 
output, constraining for the first time the geometric orientation of these sources and hence providing a more
robust spectral fit. Nevertheless, both the \citet{fernandez2007} and \citet{nobili2008a} studies make
reasonable but
heuristic assumptions about the magnetospheric physics, caused largely by the
difficulty of modeling the closed field line circuit with radiation feedback from first principles. 
 
Barring an improved understanding of the physics, 
further progress is possible by making use of
information encoded in the polarization signal. In recent years, photoelectric 
polarimeters have reached the sensitivity level required for detecting polarized X-rays from a variety of 
astrophysical sources \citep{costa2001,black2007}, allowing extension of the discovery space beyond
the lone data point for the Crab nebula \citep{novick1972}. In particular, the Gravity
and Extreme Magnetism (GEMS) mission will reach a Minimum Detectable Polarization
(MDP) of $\sim 1\%$ for a 1~mCrab source in the $2-10$~keV range with a $10^6$~s integration \citep{swank2010}. 
About one half of the current magnetar candidates
are brighter than 1~mCrab\footnote{A catalog of current magnetar candidates is maintained at
http://www.physics.mcgill.ca/$\sim$pulsar/magnetar/main.html}, making detection of their polarization a concrete possibility.

Polarized X-ray transfer calculations have been carried out for isolated neutron stars with dipole-type 
magnetic fields, in some cases including realistic atmospheres but with no magnetospheric processes other 
than propagation in a Schwarzschild spacetime
\citep{heyl2000,pavlov2000,heyl2003,vanadelsberg2006,wang2007,vanadelsberg2009,wang2009}. Both the \citet{fernandez2007}
and \citet{nobili2008a} Monte Carlo studies assume adiabatic propagation, with photons remaining in a fixed polarization eigenmode
in between scatterings, until they escape the magnetosphere. \citet{nobili2008a} estimated the observed polarization
fractions through the relative difference between the emerging normal mode intensities.
The polarization measured by a distant observer, however, is set by a gradual coupling of the propagation
eigenmodes as photons leave the magnetosphere,
with the generation of a circular
component in some cases \citep{heyl2000}. Thus, evolution of all polarization degrees of freedom is
required for making observational predictions.

In this paper we extend the multidimensional Monte Carlo radiative transfer code of \citet{fernandez2007}
to include polarization propagation and general relativistic light deflection close to the neutron star. Our focus is to
understand the basic features imparted to the polarization signal by the scattering process and the geometry
of the system. To this end, we perform radiative transfer simulations using
the simplest possible prescriptions for magnetic field geometry, particle
energy distribution, and seed photons, when the dielectric properties of the medium are dominated
by vacuum polarization. 
Motivated by GEMS, we explore the parameter dependencies of polarization
fraction and polarization angle

The structure of this paper is the following. Section 2 lays out the physics included in
the model, while Section 3 discusses our numerical implementation. Section 4 contains our 
results, focusing on the dependence of observables on system parameters. 
In Section 5 we discuss our findings in light of the expected instrumental parameters of GEMS, 
and estimate integration times for a few bright magnetar candidates. A summary and discussion 
of our results is provided in Section 6; readers not interested in technical details can 
start here, where explicit reference is made to the key figures and equations.
Appendix A contains details about geometric transformations employed in the code.

\section{Physical Model}

The elements influencing the polarization signal measured at infinity are: (i) the physics of the
magnetosphere, including the transfer of polarized radiation, (ii) the properties of the seed photons, and 
(iii) the type of spacetime. We describe our treatment of each of these items in turn. 
A detailed description of the physics entering the twisted magnetosphere model
can be found in \citet{thompson2002} and \citet{fernandez2007}. Here we provide a minimal account,
focused on the part relevant to polarization observables. 

\subsection{Twisted Magnetosphere}
\label{s:twisted_magnetosphere}

The basic hypothesis behind the magnetar model is the existence of a strong toroidal
field inside the star, thought to have been generated via dynamo action in a rapidly
rotating protoneutron star \citep{thompson1995,thompson2001a}. Seismic instabilities in the crust induced by
strong magnetic stresses mediate the transfer of some of this internal toroidal energy to the
magnetosphere, imparting a finite twist to the otherwise dipolar external field 
\citep{thompson2001a,thompson2002,beloborodov2007b}. 

To describe such a magnetosphere, one would ideally solve a closed field line circuit including
pair creation processes and feedback from radiation at cyclotron frequencies (e.g., 
\citealt{thompson2002,beloborodov2007b,thompson2008a,thompson2008b}). The global problem has not yet 
been solved self-consistently. 
Numerical computations of the X-ray emission
rely instead on prescriptions for the field geometry and particle energy distribution that
can capture the basic aspects of the model \citep{fernandez2007,nobili2008a,pavan2009}.
As our goal here is to explore the dependence of the polarization signal on different parameters
rather than to fit individual sources, we adopt the same parametric approach in our calculations.

\subsubsection{Magnetic Field Geometry and Particle Energy Distribution}
\label{s:magnetosphere}

\begin{figure}
\includegraphics*[width=\columnwidth]{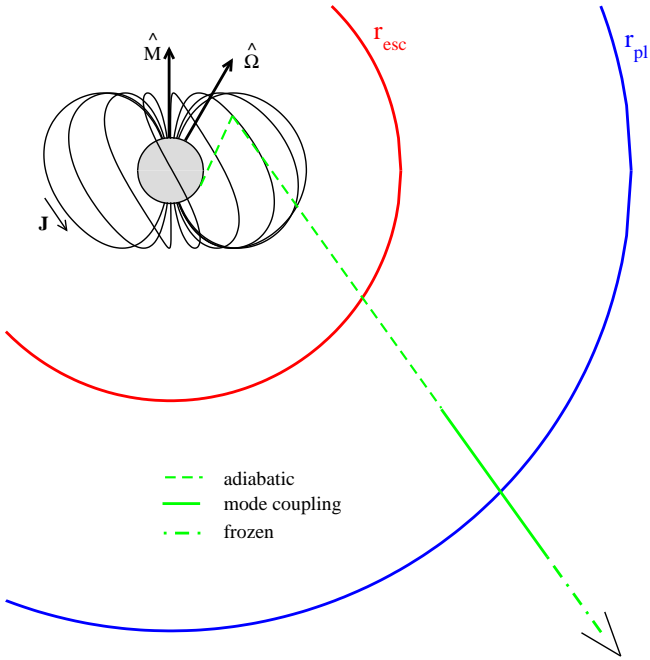} 
\caption{Diagram showing the spatial arrangement of the problem (not to scale). The neutron star (grey circle) emits
thermal photons (green trajectory) which scatter electrons moving along field lines
(black solid lines). Magnetospheric charges are provided by a current $\mathbf{J}$ due to the net magnetospheric
twist (\S\ref{s:magnetosphere}). Photons can
scatter only inside $r_{\rm esc}$ (eq.~[\ref{eq:r_escape}]). At radii much smaller than $r_{\rm pl}$ (eq.~[\ref{eq:r_pl}]), the polarization
propagation is adiabatic (dashed green line, \S\ref{s:transfer}). When approaching this radius, polarization eigenmodes start
to couple (solid green) and eventually freeze beyond $r_{\rm pl}$ (dot-dashed green). For clarity, only a few field lines anchored at
$15^\circ$ from the magnetic axis are shown, for a twist angle $\Delta \phi_{\rm N-S}=1$ (eq.~[\ref{eq:twist_angle}]).
Photon trajectories are curved close to the star due to light-bending (\S\ref{s:light_bending}). The vectors $\hat M$ and
$\hat \Omega$ denote the magnetic and rotation axes, respectively.}
\label{f:twist_diagram}
\end{figure}

We adopt the force-free, self-similar, twisted-dipole solution of \citet{thompson2002},
\begin{equation}
\label{eq:bfield_def}
\mathbf{B} = \frac{1}{2}B_{\rm pole}\left( \frac{R_{\rm NS}}{r}\right)^{2+p} \mathbf{F}(\cos\theta),
\end{equation}
with $r$ the radial distance from the center of the star, $\theta$ the polar
angle relative to the magnetic axis, $B_{\rm pole}$ the strength of the field at the poles, 
$R_{\rm NS}$ the stellar radius, and $p$ a constant. The components of the function $\mathbf{F}$ are
$F_r = -f'$, $F_\theta = (p/\sin\theta)f$, and $F_\phi = [C/\{p(p+1)\}]f^{1/p}F_\theta$,
with $f(\cos\theta)$ the solution to the second order differential equation  
\begin{equation}
\sin^2\theta\, f'' + C f^{1+2/p} + p(p+1)f = 0,
\end{equation}
subject to the boundary conditions $f'(0) = 0$, $f'(1) = -2$, and $f(1) = 0$. This leaves
an additional free parameter, either $C$ or $p$.
Alternatively, one can label solutions by the net twist angle of the field lines that are anchored
close to the magnetic poles,
\begin{equation}
\label{eq:twist_angle}
\Delta\phi_{N-S} = 2\lim_{\theta_0\to 0}\int_{\theta_0}^{\pi/2}\frac{B_\phi}{B_\theta}\frac{d\theta}{\sin\theta}.
\end{equation}
When going from $\Delta\phi_{N-S} = 0$ to $\pi$, the solutions interpolate smoothly between
a dipole ($p=1$) and a split monopole ($p=0$). 
For illustration, Figure~\ref{f:twist_diagram} shows a few field lines anchored at $15^\circ$ from
the magnetic axis for a twist $\Delta\phi_{\rm N-S} = 1$.

The current density induced along the twisted lines is \citep{thompson2002}
\begin{equation}\label{eq:current}
\mathbf{J} = \sum_i Z_ie\,n_i\, \bar\beta_i\hat{B}
c = \frac{(p+1)c}{4\pi r}\frac{B_\phi}{B_\theta}\, \mathbf{B},
\end{equation}
where $Z_ie$ is the electric charge of species $i$,
$n_i$ the number density, $\hat B = \mathbf{B}/B$, and
\begin{equation}
\bar\beta_i = \int f_i(\mathbf{r},p)\, \frac{pc}{E}\,\totd p
\end{equation}
the mean velocity along the field lines in units of $c$, with $f(\mathbf{r},p)$ the one-dimensional
phase-space distribution function, normalized such that $\int f(\mathbf{r},p)\,\totd p = 1$. 
The number density of charges $n_i$ is obtained from
\begin{equation}
\label{eq:current_sub}
\frac{|Z_i| e n_i}{B} =
\epsilon_i\frac{(p+1)}{4\pi r \bar\beta_i}\frac{B_\phi}{B_\theta},
\end{equation}
where $\epsilon_i$ is the fraction of the current carried by species $i$.

To complete the description of the magnetosphere one needs to specify the particle energy distribution
of the charge carriers. 
We employ the simplest possible prescription, a power law in momentum, independent of position,
\begin{equation}
\label{eq:pl_distribution}
f(\gamma \beta) \propto (\gamma \beta)^{-\alpha},
\end{equation}
where $\gamma = (1-\beta^2)^{-1/2}$ is the Lorentz factor. This distribution is characterized by 
three parameters: the exponent 
$\alpha$, the minimum velocity $\beta_{\rm min}$, and the maximum Lorentz factor $\gamma_{\rm max}$. In
all of our runs we employ $\alpha = -2$ and $\beta_{\rm min}=0.2$, but allow $\gamma_{\rm max}$ to
vary (see, e.g., \citealt{fernandez2007} for the resulting spectra
and their comparison with other distribution functions). 
To account for both an electron-ion and electron-positron
plasma in the magnetosphere, we allow equation~(\ref{eq:pl_distribution}) to be uni- or bi-directional, that is,
extending over only positive (from north to south magnetic pole, Fig.~[\ref{f:twist_diagram}]) or both positive
and negative momenta\footnote{In the case of an electron-ion plasma, the relevant 
charge carrier for the interaction with keV X-rays far from the star are only the electrons.}. 

In the simple case considered here ($n_e \sim J/[ec]$), the dielectric
and inverse permeability tensors are dominated by vacuum polarization
(\S\ref{s:transfer}).  As a first step in addressing this problem, and
to ease comparison with previous work, we ignore the possibility of a
large pair multiplicity, in which case $n_e \gg J/(ec)$. Such an
effect could result from pair cascades originated by the energy
deposition of current-driven instabilities and the rapid conversion of
resonantly upscattered X-rays into pairs under a strong magnetic field
\citep{thompson2008a}.  This could significantly modify the dielectric
properties at distances from the star where the adiabatic
approximation breaks down.  Given that our modeling of the
magnetospheric currents is rather heuristic, the inclusion of plasma
polarization could lead to artificial features sensitive to the
temperature and density of the magnetospheric particles.  Hence, we
opt for a self-consistent but potentially incomplete calculation, and
only include vacuum polarization.  Consequently, we set $\epsilon_i
\equiv \epsilon = 1$ in equation~(\ref{eq:current_sub}) for an
ion-electron plasma, or $\epsilon_i = 1/2$ when electrons and
positrons are present.

\subsubsection{Resonant Cyclotron Scattering}
\label{s:scattering}

For magnetar field strengths, the lifetime of an excited
Landau level is very short, $ \Delta t_{\rm L} = (3/4)(\hbar/[\alpha_{em}\,m_e c^2])(B_{\rm QED}/B)^2 
\simeq 3\times 10^{-14}B_{11}^{-2}$~s, with $\alpha_{\rm em}$ the fine structure constant, 
$B_{\rm QED} = m_e^2 c^3/(\hbar e) \simeq 4.4\times 10^{13}$~G the field strength where the
cyclotron energy equals the electron rest mass energy, and $B_{11} \equiv B/(10^{11}{\rm G} )$.
Hence, electrons or positrons can be assumed to
be in the ground state, and absorption plus re-emission of photons can be treated as
a single scattering process \citep{zheleznyakov96}. Excitation occurs when the photon
frequency, in the rest frame of the charge, equals the cyclotron frequency.
In the stellar frame, this condition corresponds to
\begin{equation}
\label{eq:omega_doppler}
\omega = \omega_\mathrm{D} \equiv \frac{\omega_c}{\gamma (1-\beta \mu)},
\end{equation}
where $\omega$ is the photon angular frequency, 
\begin{equation}
\mu = \hat k\cdot \hat B
\end{equation}
is the cosine of the angle between photon ($\hat k$) and magnetic field directions in the 
\emph{stellar frame}, and 
\begin{equation}
\omega_c = \frac{eB}{m_e\,c} \simeq 1.2B_{11}~\frac{{\rm keV}}{\hbar}
\end{equation} 
the electron cyclotron frequency. 

The total cross section in the stellar frame is given by (e.g., \citealt{meszaros1992})
\begin{equation}
\label{eq:cross_section}
\sigma_\mathrm{res} = 4\pi^2\, (1-\beta\mu)\,\frac{|Z|e}{B}|e_\pm|^2\,
\omega_\mathrm{D}\delta(\omega-\omega_\mathrm{D}),
\end{equation} 
where $|e_\pm|$ is the \emph{rest frame} overlap of the photon polarization state with a left (-) or right (+)
circularly polarized wave, depending on the sign of the charge. The factor
$(1-\beta\mu)$ comes from the relativistic boost to the cross section (e.g., \citealt{rybicki2004}). 
The delta function is an approximation to the small fractional width of the resonance for a single particle, 
$(2\Delta t_{\rm L}\omega_c)^{-1} = (2\alpha_{\rm em}/3) (B/B_{\rm QED}) = 10^{-5}B_{11}$. The enhancement
relative to the usual non-resonant Thomson cross section $\sigma_T$ is, at resonance and in the charge rest frame,  
$\sigma_{\rm res}/\sigma_T = |e_\pm|^2 (2\Delta t_{\rm L}\omega_c)^2 \sim 10^{10}\,|e_\pm|^2\,B_{11}^2$ \citep{meszaros1992}. 

Calculation of the optical depth requires averaging the cross section over the phase space
distribution of the scattering charges. Using the fact that for fixed photon frequency
and at a given position, equation~(\ref{eq:omega_doppler}) has two roots $\beta^{\pm}$,
the differential optical depth along a small distance $\Delta \ell \ll r$ is given by \citep{fernandez2007} 
\begin{eqnarray}\label{eq:dtau}
\Delta \bar \tau_{\rm res} = & & \varepsilon\frac{\pi(p+1)\omega}{r|\bar\beta|}
\left(\frac{B_\phi}{B_\theta}\right)\left[(1-\beta^+\mu)|e_\pm|^2_{\gamma\beta^+} \frac{f(\gamma\beta^+)\,\Delta(\gamma\beta)^+}
{\left(\partial \omega_{D}/\partial \ell\right)_{\gamma\beta^+}}\right.\nonumber\\
& & \left.- (1-\beta^-\mu)|e_\pm|^2_{\gamma\beta^-}\frac{f(\gamma\beta^-)\,\Delta (\gamma\beta)^-}
{\left(\partial \omega_\mathrm{D}/\partial\ell\right)_{\gamma\beta^-}}\right],
\end{eqnarray}
where $(\gamma \beta)^{\pm}$ are the resonant momenta corresponding to the resonant velocities $\beta^{\pm}$,
$\Delta(\gamma\beta)^\pm$ is the size of the step in resonant-momentum space corresponding to the spatial step $\Delta \ell$,
and $\partial \omega_D/\partial \ell = \hat k \cdot \nabla \omega_D$.
Equation~(\ref{eq:dtau}) differs from \citet{fernandez2007} in that the average velocity $|\bar \beta|$
comes out of the integral, as it is given by the relation between current density and number density
(eq.~[\ref{eq:current}]) and does not depend on the resonance condition\footnote{\citet{nobili2008a} used 
an optical depth with this error corrected. The factor $\omega_{\rm D}^{-1}(\partial \omega_{\rm D}/\partial \ell)$ 
is independent of $\omega_D=\omega$, so there is no dependence on photon frequency or magnetic field strength.}.
Given that typical values of $\beta$ are not too far from the mean of the distribution, this modification does not introduce a 
significant difference in the results relative to those of \citet{fernandez2007}.

During scattering, a photon can change its polarization state. The probabilities depend on the rest frame overlap of 
the outgoing photon polarization with a circularly polarized wave, as the differential cross section for scattering from 
polarization state $A$ to state $B$ satisfies \citep{meszaros1992}
\begin{equation}
\label{eq:diff_cross_sec}
\frac{\totd \sigma_{\rm res}}{\totd \Omega^\prime} \propto |e_\pm|_A^2 |e^\prime_\pm|_B^2,
\end{equation}
where primes denote outgoing photon quantities.
If normal modes begin to couple, the simple adiabatic expressions for the \emph{ingoing} photon overlap need to be modified,
as described in \S\ref{s:numeric_scattering} (eq.~[\ref{eq:non_adiabatic_overlap}]). 
For the \emph{outgoing} photon, however, we impose a normal mode
state, regardless of the degree of adiabaticity. As vacuum polarization dominates the dielectric properties (\S\ref{s:transfer}),
the electric field eigenvectors are given by $\hat e_E \propto (\hat k \times \hat B)$ and $\hat e_O \propto (\hat e_E \times \hat k)$.
The overlap is the dot product of the normalized eigenvectors with $\hat e_\pm = \left(\hat x \pm i\hat y\right)/\sqrt{2}$ in a 
frame where $\hat B$ is along $\hat z$, yielding
\begin{equation}
\label{eq:overlaps_outgoing}
|e^\prime_\pm|_E^2 = 1/2, \qquad |e^\prime_\pm|_O^2 = (\mu_r^\prime)^2/2,
\end{equation}
where
\begin{equation}
\label{eq:murest}
\mu_r = \frac{\mu - \beta}{1 - \beta\mu}
\end{equation}
is the angle between photon direction and magnetic field in the \emph{rest frame} of the charge.
The probability of the photon coming out in E-mode is then
\begin{equation}
\label{eq:mode_switch_prob}
P(E) = \frac{1}{1 + (\mu_r^\prime)^2},
\end{equation}
with $P(O) = 1 - P(E)$. 
In practice, imposing a normal mode right after scattering does not affect our results, as the polarization coupling and 
scattering regions are well separated in space (\S\ref{s:transfer}, see also Figure~\ref{f:surfaces}).

Note that for all $|\mu_r^\prime|<1$, $P(O) < P(E)$, and hence E-mode photons always dominate
the outgoing polarization distribution except for exactly parallel propagation.
 
\subsubsection{Transfer of Polarized Radiation}
\label{s:transfer}

We first analyze the dielectric properties of the magnetosphere, then write down the
polarization evolution equations, and finally characterize the regions over which magnetospheric 
scattering and non-adiabatic propagation are relevant, looking for possible overlaps.

For $B \ll B_{\rm QED}$, the relative contributions of plasma and vacuum polarization to the dielectric
and inverse permeability tensors can be quantified by the ratio (e.g., \citealt{zheleznyakov96})
\begin{eqnarray}
\label{eq:dielectric_ratio}
\frac{(\omega_p/\omega)^2/|\bar\beta|}{\alpha_{\rm em}(B/B_{\rm QED})^2/45\pi} \simeq 
10^{-7}\,B_{{\rm pole},14}^{-1} R_6^{-1}\left( \frac{r}{R_{\rm NS}}\right)^{1+p}\nonumber \\
\times\frac{\sin^2\theta\,\Delta\phi_{\rm N-S}}{|\bar\beta|} \left(\frac{\hbar\omega}{\rm keV}\right)^{-2},
\end{eqnarray}
where $\omega_p = \sqrt{4\pi\,Z^2e^2\,n_e/m_e}$ is the electron plasma frequency, 
and the current density (eq.~[\ref{eq:current}]) has been approximated by
$\mathbf{J} \simeq c(4\pi\, r)^{-1}\sin^2\theta\,\Delta\phi_{\rm N-S}\,\mathbf{B}$ \citep{thompson2002}.
Equation~(\ref{eq:dielectric_ratio}) shows that vacuum polarization dominates out to $r\simeq 3000R_{\rm NS}$
for mildly relativistic particle distributions ($|\bar\beta|\lesssim 1$). Outside this radius plasma polarization
starts to dominate.

We show in \S\ref{s:numeric_polarization} that the polarization vector effectively freezes at distances 
$\lesssim 500R_{\rm NS}$, well inside the vacuum dominated regime. However, this would change if the pair 
multiplicity is bigger than $\mathcal{M}_\pm \sim 10^{3}$, because the radius where plasma and 
vacuum polarization become comparable would decrease by a factor $(\mathcal{M}_\pm)^{1/(1+p)}\sim 30$.

Due to quantum electrodynamic effects, a strong magnetic field modifies the dielectric properties of 
the vacuum according to \citep{klein1964,adler1971}
\begin{eqnarray}
\label{eq:dielectric_tensor}
\varepsilon & = & (1+a)\,\mathbb{I} + q\,\hat{B}\hat{B}\\
\label{eq:permeability_tensor}
\bar{\mu}   & = & (1+a)\,\mathbb{I} + m\,\hat{B}\hat{B},
\end{eqnarray}
where $\varepsilon$ and $\bar\mu$ are the dielectric and inverse permeability tensors, respectively,
and $\mathbb{I}$ is the unit tensor. For $B\ll B_{\rm QED}$, the coefficients are $a = -2\delta$, 
$q = 7\delta$, and $m = -4\delta$, with
$\delta = \alpha_{\rm em}(45\pi)^{-1}(B/B_{\rm QED})^2\simeq 3\times 10^{-10}B_{11}^2$. These coefficients
are all real, so there is no change in the total intensity.
The propagation 
eigenmodes are linear, and are defined in relation to the photon and magnetic field directions (\S\ref{s:scattering}). 
General expressions exist for arbitrary field strengths, which
nevertheless maintain the geometric structure of equations~(\ref{eq:dielectric_tensor})-(\ref{eq:permeability_tensor}) [\citealt{heyl1997}].
Given that polarization eigenmodes begin to decouple at distances $r\gtrsim 10R_{\rm NS}$, the weak field
limit is good enough for our purposes.

The polarization of photons can be characterized through the complex components of the electric
field vector in some coordinate system, or through the Stokes parameters (e.g., \citealt{rybicki2004}).
In our implementation, we choose to evolve the former, as it is simpler to perform geometric transformations on them,
and convert to Stokes parameters only after the polarization is deemed to be frozen (\ref{s:numeric_polarization}).

We now derive the ordinary differential equations governing the
evolution of the electric field vector in the geometric optics limit,
following the notation of \citet{lai2003b}.
The wave equation for the electric field of a plane electromagnetic wave $\mathbf{E}= \mathbf{E}_0(\mathbf{r}) e^{-i\omega t}$ is
\begin{equation}
\label{eq:maxwell_wave}
\nabla \times \left[\bar{\mu}\cdot (\nabla \times \mathbf{E}) \right] = \frac{\omega^2}{c^2}\varepsilon\cdot \mathbf{E}.
\end{equation}
The polarization tensors are assumed to be constant in time and varying over distances much larger than the photon wavelength.
We then choose a coordinate system with the $z$-axis along the direction of photon propagation,
assume that quantities change only along $z$, and write $\mathbf{E}_0(z) = \exp\left(ik_0z\right)\mathbf{A}(z)$, 
with $k_0 = \omega/c$. The geometric optics approximation is equivalent to $\partial_z \mathbf{A} \ll k_0 \mathbf{A}$.
Keeping terms linear in $k_0^{-1}\partial_z$, we obtain a system of ordinary differential equations describing the propagation
of the complex amplitude $\mathbf{A}$,
\begin{eqnarray}
\label{eq:prop_ax_general}
\partial_z A_x & = & \frac{ik_0}{2}\,\left(\bar{\mu}_{yy} - \frac{\bar{\mu}_{xy}\bar{\mu}_{yx}}{\bar{\mu}_{xx}} \right)^{-1}\,
 \left[ \left( \varepsilon_{xx}-\bar{\mu}_{yy} + \frac{\bar{\mu}_{yx}}{\bar{\mu}_{xx}}\left[\varepsilon_{yx}+\bar{\mu}_{xy}\right]\right)\,A_x
     \right.\nonumber\\ 
& &+\left.\left( \varepsilon_{xy}+\bar{\mu}_{yx} + \frac{\bar{\mu}_{yx}}{\bar{\mu}_{xx}}\left[\varepsilon_{yy}-\bar{\mu}_{xx}\right]\right)\,A_y
     \right]\\
\label{eq:prop_ay_general}
\partial_z A_y & = & \frac{ik_0}{2}\,\left(\bar{\mu}_{xx} - \frac{\bar{\mu}_{xy}\bar{\mu}_{yx}}{\bar{\mu}_{yy}} \right)^{-1}\,
 \left[ \left( \varepsilon_{xy}+\bar{\mu}_{yx} + \frac{\bar{\mu}_{xy}}{\bar{\mu}_{yy}}\left[\varepsilon_{xx}-\bar{\mu}_{yy}\right]\right)\,A_x 
    \right.\nonumber\\
& &+\left.\left( \varepsilon_{yy}-\bar{\mu}_{xx} + \frac{\bar{\mu}_{yx}}{\bar{\mu}_{yy}}\left[\varepsilon_{xy}+\bar{\mu}_{yx}\right]\right)\,A_y
    \right]\\
\label{eq:prop_az_general}
\phantom{\partial_z}A_z & = & -\frac{\varepsilon_{zx}}{\varepsilon_{zz}}\,A_x - \frac{\varepsilon_{zy}}{\varepsilon_{zz}}\,A_y.
\end{eqnarray}
Equations~(\ref{eq:prop_ax_general})-(\ref{eq:prop_az_general}) are valid for general conditions, so long as the geometric
optics and smooth background approximation holds. Inclusion of the plasma contribution to $\bar\mu$ and $\varepsilon$ is 
mathematically straightforward, once the number density and temperature of charges as a function of position are known (see, e.g.,
\citealt{lai2003b} for an application to neutron star atmospheres).

Even for magnetar field strengths, the non-diagonal components of $\varepsilon$ are much smaller that unity, hence
from equation~(\ref{eq:prop_az_general}) we can neglect $A_z$. 
To more clearly visualize the behavior of equations~(\ref{eq:prop_ax_general})-(\ref{eq:prop_ay_general}), 
one can switch to a reference frame where the magnetic field is in the $x$-$z$ plane, 
$\hat{B} = \cos\theta_{kB}\hat{k} + \sin\theta_{kB}\hat{x}$, and keep terms linear in $\{a,q,m\}$, finding
\begin{equation}
\label{eq:prop_diagonal}
\frac{\partial}{\partial z}\left( \begin{array}{c} A_x\\A_y\end{array}\right) = \frac{i}{2}k_0 \sin^2\theta_{kB}\,
\left( \begin{array}{cc} q & 0\\ 0 & -m \end{array}\right)\,\left( \begin{array}{c} A_x\\ A_y\end{array}\right).
\end{equation}
In this case, the $x$ and $y$ components correspond to the O- and E-mode eigenvectors, respectively.
The characteristic scale over which $\mathbf{A}$ varies is 
\begin{equation}
\label{eq:l_A_def}
\ell_A = \frac{1}{k_0\,\Delta n},
\end{equation}
where $\Delta n = \sin^2\theta_{kB}(q+m)/2$ is the difference between the indices of refraction \citep{heyl2000,lai2003b}. 
As the photon propagates, non-diagonal components will appear in equation~(\ref{eq:prop_diagonal}),
because $\hat{B}$ will leave the $x-z$ plane. However, if this occurs over a length-scale much larger than $\ell_A$, normal
modes do not mix, and the propagation is \emph{adiabatic} \citep{heyl2000,lai2003b}. As the photon moves out in the magnetosphere,
the QED correction to the dielectric tensor falls off as $B^2\sim r^{-6}$, causing $\ell_A$ to increase in magnitude,
until the point at which the polarization effectively freezes. The distance from the star at which the adiabatic approximation
breaks down is usually called the \emph{polarization limiting radius} \citep{heyl2003}, and is given by the radius at
which the relation $B/|\hat{k}\cdot \nabla B|\sim \ell_A$ is satisfied
(see \citealt{heyl2003,lai2003b} for slightly different criteria).
The coupling of the modes is gradual, however, and a characterization of the observed polarization 
signal requires solution of equations~(\ref{eq:prop_ax_general})-(\ref{eq:prop_ay_general}) over this coupling region.

It is instructive to quantify the size of the regions where resonant scattering 
and mode coupling are relevant, as this determines the simplifications that are possible 
in the full problem. The surface inside which resonant cyclotron scattering can occur is the
\emph{escape radius}, given by the condition $(\omega_c/\omega)^2 + \mu^2 = 1$ \citep{fernandez2007}.
Outside of this surface, there is no particle velocity such that the resonance condition 
(eq.~[\ref{eq:omega_doppler}]) can be satisfied.
For the twisted dipole geometry, it is given by
\begin{eqnarray}
\label{eq:r_escape}
r_{\rm esc} & = & r_{\rm esc}(\theta,\omega,\hat k,B_{\rm pole},\Delta \phi_{N-S},R_{\rm NS}) \nonumber\\
\label{eq:r_esc}
     & = & \left[ \xi\,\left( 1-\mu^2\right)^{-1/2}\,\left(\frac{\omega_{\rm c,pole}}{\omega}\right)
\right]^{1/(2+p)}\, R_{\rm NS}\\
     & \simeq & 12\left[ \xi\,\left( 1-\mu^2\right)^{-1/2}B_{\rm pole, 14} 
\left(\frac{\rm keV}{\hbar \omega}\right)\right]_{\Delta\phi=1}^{1/2.88}\,R_{\rm NS}\nonumber,
\end{eqnarray}
where $\omega_{\rm c,pole} = eB_{\rm pole}/(m_e c)$, and $2\xi = [\mathbf{F}^2(\cos\theta)]^{1/2}$ is
the angular dependence of the magnetic field strength (c.f. eq.~[\ref{eq:bfield_def}]).
The polarization limiting radius is set by the equality between $\ell_A$ 
and the length scale over which the field strength changes, $\ell_B = B/|\hat k \cdot \nabla B|$.
Writing $\ell_B = r/\zeta(\theta,\hat k,\Delta\phi_{\rm N-S})$, where $\zeta$ is a dimensionless
function of order unity, we find
\begin{eqnarray}
r_{\rm pl} & = & r_{\rm pl}(\theta,\omega,\hat k, B_{\rm pole}, \Delta \phi_{\rm N-S},R_{\rm NS})\nonumber \\
\label{eq:r_pl}
           & = & R_{\rm NS}\,\left[ \left(\frac{\alpha_{\rm em}}{30\pi}\right)\,
\left(\frac{B_{\rm pole}}{B_{\rm QED}}\right)^2 (1-\mu^2) \frac{\xi^2}{\zeta} \left(\,\frac{R_{\rm NS}\,\omega}{c}\right)
\right]^{1/(3+2p)}\\
           & \simeq & 146 \left[(1-\mu^2)\frac{\xi^2}{\zeta}B_{\rm pole, 14}^2 R_6
             \left(\frac{\hbar\omega}{\rm keV}\right)\right]_{\Delta\phi=1}^{1/4.76}\,R_{\rm NS},
\end{eqnarray}
where $R_6 = R_{\rm NS}/10^6$~cm.

\begin{figure}
\includegraphics*[width=\columnwidth]{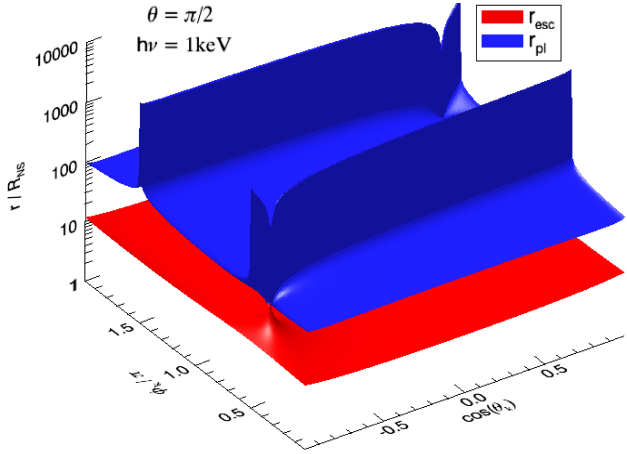}
\caption{Escape radius $r_{\rm esc}$ (eq.~[\ref{eq:r_esc}], red surface) and polarization limiting radius
$r_{\rm pl}$ (eq.~[\ref{eq:r_pl}], blue surface) for $1$~keV photons at the magnetic equator, as a function of photon direction
$\hat k(\theta_k, \phi_k)$, with $\theta_k$ and $\phi_k$ the polar and azimuthal angle of $\hat k$ relative
to the magnetic axis, respectively. 
Other parameters are $B_{\rm pole} = 10^{14}$~G, $\Delta \phi_{\rm N-S} = 1$~rad, and $R_{\rm NS} = 10^6$~cm. 
The spikes in the surfaces arise at points where the expressions become singular, namely $|\mu| \to 1$ for $r_{\rm esc}$ 
and $\hat k \cdot \nabla B \to 0$ for $r_{\rm pl}$. 
}
\label{f:surfaces}
\end{figure}

Figure~\ref{f:surfaces} shows $r_{\rm esc}$ and $r_{\rm pl}$ as a function of photon direction angles, 
for $1$~keV photons at the magnetic equator, with a twist $\Delta \phi_{\rm NS} = 1$ and a standard choice for other parameters. 
Aside from narrow regions where the surfaces become singular ($|\mu|\to 1$ for $r_{\rm esc}$ and $\hat k \cdot \nabla B\to 0$
for $r_{\rm pl}$), they are well separated by an order of magnitude in radius. Photons in the Rayleigh-Jeans tail
of the thermal spectrum could in principle violate this separation, but they contribute little to the observed signal.
One can thus assume that (1) scattering is largely decoupled from polarization
evolution, taking place mostly within the adiabatic propagation region, and (2) the polarization signal is set
at distances $\gtrsim 100R_{\rm NS}$ from the center, where the effects of general relativity are 
weak. 
\newline

\subsection{Seed Photon Distribution}
\label{s:seed_distribution}

Two physical processes have been proposed to explain the thermal component of
the quiescent soft X-ray spectrum. The first is heat generated in the
core by ambipolar diffusion \citep{thompson1996}. These photons are expected
to emerge in E-mode polarization, as their free-free opacity is suppressed
relative to the O-mode (e.g., \citealt{lai2002}). Including resonant mode conversion
at magnetar field strengths ($\gtrsim 10^{14}$~G at the surface) does not
alter this dominance of the E-mode, but instead modifies the spectral shape (e.g., \citealt{lai2003a}).
For the purposes of exploring 
the effects of magnetospheric scattering on the surface polarization, we adopt a simple
blackbody spectrum with a temperature such that its value measured at infinity
is $kT = 0.4$~keV. This case is assigned a $100\%$ initial polarization in the E-mode.

The second source of thermal photons is the deposition of energy at the stellar
surface by \emph{returning currents} in a layer that is optically thick to 
free-free absorption \citep{thompson2002,beloborodov2007b}. In this case, the emission is
expected to emerge predominantly in O-mode polarization. Since the details of
this mechanism have not been thoroughly explored yet, we account for this possibility by 
simply changing the outgoing polarization distribution from E- to O-mode.
Intermediate mixtures (such as a completely unpolarized case) can be obtained
by linear superposition.

Regardless of the mechanism responsible for the seed photons, emission is expected
to be anisotropic over the stellar surface. For the deep cooling case, even 
a constant temperature atmosphere has an emerging radiation intensity that is beamed 
relative to the magnetic field direction, with an angular distribution that depends on 
photon energy and magnetic field strength \citep{vanadelsberg2006}. Additional 
anisotropies are expected due to the suppression of thermal conduction across
magnetic field lines. If a strong internal toroidal field is present, very narrow
polar caps are expected \citep{perez-azorin2006,geppert2006}.
If the thermal emission is powered by returning currents, localized hot spots
of the size of the current carrying region would be natural.
We explore the effects of anisotropies in the surface temperature
by running additional models restricting emission to polar caps with a size of 5 degrees. 
We do not attempt here to include the effect of beaming of radiation around the local
magnetic field direction. It is worth keeping in mind, though, that general relativistic 
deflection of photon trajectories significantly smears out emission from a polar cap, 
enlarging its apparent angular size (\S\ref{s:light_bending}).

\subsection{Spacetime Properties}
\label{s:light_bending}

Magnetars are slow rotators, with periods falling in the range $2-10$~s (e.g., \citealt{woods2006}). At the surface of the star,
corrections to the metric due to rotation are of order $(J_{\rm NS}/[M_{\rm NS}\,R_{\rm NS}\,c])^2\sim
\left(R_{\rm NS}\Omega/c\right)^2\lesssim 4\times 10^{-8}\,R_6^2\,P_{\rm sec}^{-2}$, where $J_{\rm NS}$, 
$M_{\rm NS}$, and $\Omega = 2\pi/P$ are the stellar angular momentum, mass, and angular
velocity, respectively, and $P_{\rm sec} = P/1\textrm{ s}$. Hence,
using a Scharzschild spacetime is a very good approximation.

In \S\ref{s:transfer} we showed that the polarization signal is determined at radii
$\gtrsim 100R_{\rm NS}$, where the effects of general relativity on photon propagation
are weak. However, a significant fraction of the seed photons leaves the magnetosphere
without scattering, thus inclusion of light-bending
is needed to properly account for the fraction of the stellar surface contributing to the
unscattered polarization signal. This is particularly important when including emission
from hot spots, as bending angles are as large as $30$ degrees for
typical neutron star parameters (e.g., \citealt{beloborodov2002}).

We adopt the approach of \citet{heyl2003} for inclusion of general relativity in our polarization calculation.
That is, (1) modifications to the magnetic field are ignored, as close to the star this is
of weak importance for both scattering and polarization\footnote{It would affect scattering for either very
high energy photons, very energetic particle distributions, or ion scattering. In the case of polarization transfer, propagation 
close to the star is well inside the adiabatic limit; see \S\ref{s:transfer}.};
(2) the polarization plane is parallel transported along photon geodesics; and (3) we ignore 
general relativistic corrections to the polarization transfer 
equations (\ref{eq:prop_ax_general}-\ref{eq:prop_ay_general}) other than the modification 
of the polarization plane.

\section{Numerical Implementation}
\label{s:monte_carlo}

Our Monte Carlo code has been extensively described in \citet{fernandez2007}. We describe here the modifications 
introduced in the present study.

\subsection{Propagation}

Integration of photon trajectories in a Scharzschild metric is a standard textbook calculation
(e.g., \citealt{schutz2009}). However, for practical purposes, we recast the equations in a form
more amenable for the Monte Carlo algorithm. The calculation
proceeds as follows. For each spatial step of length $\Delta \ell$, the coordinate system is rotated to a frame where
$\hat x = \hat r$ and $\hat z = \hat r \times \hat k$, e.g., the usual $\theta = \pi/2$
plane (see Appendix~\ref{s:rotations} for the explicit transformation). We
then integrate, as a function of azimuthal angle, the geodesics for the radial coordinate $r$,
the components of $\hat k$ in a non-coordinate orthonormal polar basis, 
$\hat k = k^{\hat r}\hat r + k^{\hat \phi}\hat \phi$, where $\hat r = \sqrt{(1-r_s/r)}\,\vec e_r$ and 
$\hat \phi = \vec e_\phi/r$ (e.g., \citealt{schutz2009}), and the line segment traversed $\Delta s$. 
Here and throughout the paper, $r_s \equiv 2GM_{\rm NS}/c^2$.
The ordinary differential equation system is
\begin{eqnarray}
\label{eq:light_bend_r}
\frac{\totd r}{\totd \phi}             & = & r\sqrt{1-\frac{r_s}{r}}\left(\frac{k^{\hat r}}{k^{\hat\phi}}\right)\\
\frac{\totd k^{\hat r}}{\totd \phi}    & = & \frac{k^{\hat \phi}}{\sqrt{1 - r_s/r}}\left[1 - \frac{3}{2}\frac{r_s}{r} \right]\\
\frac{\totd k^{\hat \phi}}{\totd \phi} & = & -k^{\hat r}\sqrt{1 - \frac{r_s}{r}}\left[1 - \frac{1}{2}\frac{r_s/r}{(1-r_s/r)} \right]\\
\label{eq:light_bend_s}
\frac{\totd s}{\totd \phi}             & = & \sqrt{\left(\frac{\totd r}{\totd \phi}\right)^2 + r^2}.
\end{eqnarray}
We have tested our implementation by verifying agreement with the asymptotic expansion of \citet{beloborodov2002}, which is
accurate to about 1 part in $10^{-3}$.

Within the code, the problem is of boundary value type, as we require the line segment traversed $\Delta s$ to equal the 
externally imposed step size $\Delta \ell$. This is needed so that a backwards and forward step sequence (e.g., for spatial
derivatives of the magnetic field) returns to the original position. In practice, we have found that a simple 3-step 
iterative process achieves
this equality to double precision in most cases. The initial angular interval for the Runge-Kutta integrator is estimated as 
$\Delta \phi_1 = \Delta \ell / \sqrt{r^2 + (\totd r/\totd \phi)^2}$. Subsequent corrections are then
$\Delta \phi_{i+1} = \left(\Delta \ell - \Delta s_i \right)/\sqrt{r_i^2 + (\totd r/\totd\phi)_i^2}$, where $i$ is
the order of the iteration. Despite these simplifications, combined solution of 
equations~(\ref{eq:prop_ax_general})-(\ref{eq:prop_ay_general}) 
and (\ref{eq:light_bend_r})-(\ref{eq:light_bend_s}) becomes computationally expensive when
$\gtrsim 10^7$ photons are followed. We thus cut off light bending when $r_s/r < 2\times 10^{-2}$, i.e., when general relativity
modifies the trajectory to less than 1\%, and impose planar propagation outside that radius.

\subsection{Scattering}
\label{s:numeric_scattering}

Aside from taking $\bar\beta$ out of the square brackets in 
equation~(\ref{eq:dtau}), the scattering algorithm remains the same as in \citet{fernandez2007}, 
including the adaptive step size algorithm in momentum space.

The overlap of the photon polarization with a circularly polarized wave $|e_{\pm}|^2$, which enters the optical depth (eq.~[\ref{eq:dtau}]),
needs to be modified for the possibility of non-adiabaticity, 
\begin{eqnarray}
\label{eq:non_adiabatic_overlap}
|e_\pm|^2 & = & \frac{1}{2}\bigg|\mathbf{A}\cdot\left(\mu_r\hat x - \sqrt{1-\mu_r^2}\hat z \pm i\hat y \right)\bigg|^2\nonumber \\
          & = & \frac{1}{2}\left[\mu_r{\rm Re}(A_x)\mp {\rm Im}(A_y) \right]^2\nonumber\\ 
          &   & + \frac{1}{2}\left[\mu_r{\rm Im}(A_x) \pm {\rm Re}(A_y)\right]^2,
\end{eqnarray}
where $\hat k = \hat z$, and the magnetic field lies in the $x-z$ plane.
Thus, when modes are coupled, the sign of the charge needs to be taken into account.
\vspace{0.3in}

\subsection{Polarization Evolution} 
\label{s:numeric_polarization}

Since $\ell_A \propto B^{-2}$, this length scale can become very small at radii $r \ll r_{\rm pl}$, hence integration of 
equations~(\ref{eq:prop_ax_general})-(\ref{eq:prop_ay_general}) becomes impractical in regions where
adiabatic propagation is a good approximation. An arbitrary choice has to be made for the transition
surface where the ordinary differential equation solver is turned on. Once this choice is made, 
we set the initial conditions for the integration 
such that the absolute value of the components of the electric field correspond to those of the
normal mode the photon was in before reaching the transition surface. A single random complex phase
is then assigned to all components.

\begin{figure}
\includegraphics*[width=\columnwidth]{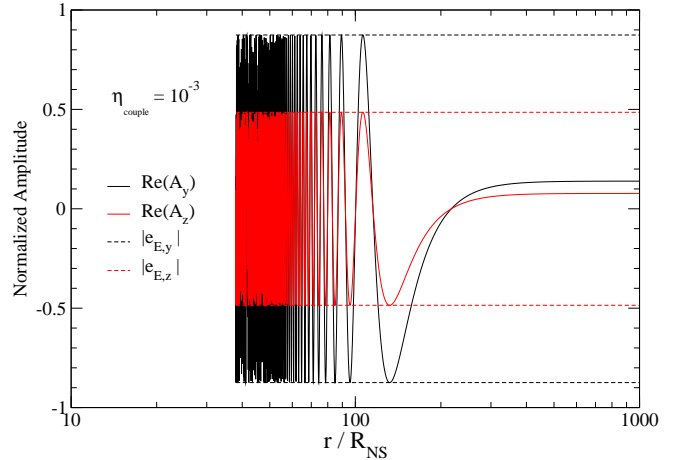}
\caption{Real components of the electric field (solid lines) and normal mode eigenvectors (dashed-lines)
for a test E-mode photon moving radially outward along the $x$-axis. Integration of 
equations~(\ref{eq:prop_ax_general})-(\ref{eq:prop_ay_general}) begins when $\eta_{\rm couple} = 10^{-3}$ 
(eq.~[\ref{eq:lcouple}]). The photon stays in the same normal mode because the component of $\mathbf{B}$
transverse to the propagation does not change direction.}
\label{f:test_radial_E}
\end{figure}

Figure~\ref{f:test_radial_E} shows the real part of the electric field vector relative to the
magnetic axis for a test E-mode photon moving outwards along the $x$-axis, in a 
magnetosphere with a twist $\Delta \phi_{\rm NS} = 1$. Integration of 
equations~(\ref{eq:prop_ax_general})-(\ref{eq:prop_ay_general}) begins at
\begin{equation}
\label{eq:lcouple}
\eta_{\rm couple} \equiv \frac{\ell_A}{r} = 10^{-3}, 
\end{equation}
corresponding to a radius $r \simeq 40R_{\rm NS}$ on the equatorial plane. As the photon travels out, 
the characteristic spatial oscillation frequency of $\mathbf A$ becomes longer, 
and the polarization reaches its asymptotic value
at a distance $\sim 200R_{\rm NS}$ from the star. In this particular case, because the trajectory is radial,
the component of the magnetic field perpendicular to the trajectory does not change direction,
thus normal modes should not couple (c.f. eq.~[\ref{eq:prop_diagonal}]).
Figure~\ref{f:test_radial_E} shows that the amplitude of the oscillations in $\mathbf A$ are
bracketed by the absolute value of the corresponding polarization eigenvector,
hence our integration correctly captures adiabatic propagation. This purity of normal
mode state is achieved to double precision for radial propagation, as measured by
the intensity in the orthogonal mode.

\begin{figure}
\includegraphics*[width=\columnwidth]{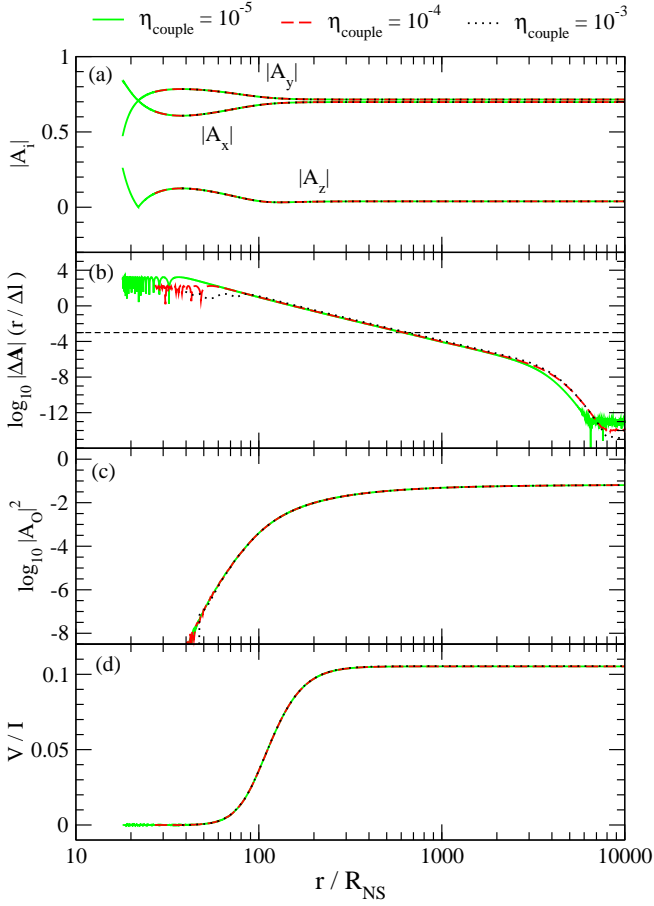}    
\caption{Various quantities for a test E-mode photon moving along the line $18R_{\rm NS}\hat x + 
\lambda (\hat x + \hat y + \hat z)$ relative to the magnetic axis, where $\lambda > 0$ is a
parameter labeling the trajectory. Different curves correspond to different values $\eta_{\rm couple}$ above which
the polarization evolution equations~(\ref{eq:prop_ax_general})-(\ref{eq:prop_ay_general}) are solved. 
Panels show, as a function of radial distance from the star, (a) the absolute value of the electric
field vector components relative to the magnetic axis, (b) the approximate change of the polarization
vector over one magnetic
field scale-height, (c) the projection of the electric field onto the O-mode
eigenvector, and (d) the normalized Stokes $V$ parameter (eq.~[\ref{eq:stokes_V}]). The dashed line shows the value of the 
electric field change at which we consider the polarization frozen (see text for details).}
\label{f:trajectory_error}
\end{figure}

\begin{figure}
\includegraphics*[width=\columnwidth]{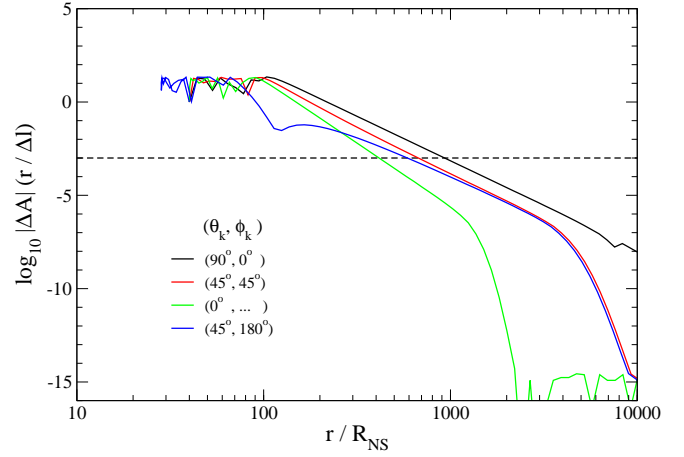}
\caption{Norm of the change in the polarization vector over a scale height in magnetic field (eq.~[\ref{eq:A_error}])
for various trajectories, labeled by the direction of the wave vector $\hat k$ relative to the magnetic axis.
The latter is parameterized in terms of the polar and azimuthal angles $\theta_k$ and $\phi_k$, respectively.
Photons start from $\mathbf r = 40 R_{\rm NS}\hat x $, and their wave vector points towards $\hat x$ (black),
$[\hat x + \hat y + \hat z]$ (red), $\hat z$ (green), and $[-\hat x + \hat z]$ (blue). The latter
trajectory crosses the $z$-axis. The horizontal dashed line marks the point where we consider
the polarization to be effectively frozen.}
\label{f:error_trajectory_k}
\end{figure}

Figure~\ref{f:trajectory_error} shows the effect of changing the value of $\eta_{\rm couple}$
for a test E-mode photon moving along $\mathbf{r} = 18R_{\rm NS}\hat x + 
\lambda(\hat x +\hat y + \hat z)$, where $\lambda > 0$ is a parameter labeling the
trajectory\footnote{We ignore light-bending in Figures~\ref{f:trajectory_error} and \ref{f:error_trajectory_k}.}.
Because adiabatic propagation is captured with good accuracy, there is no significant difference in starting  the integration
as deep as $\eta_{\rm couple} = 10^{-5}$, corresponding roughly to a factor two in distance relative to $\eta_{\rm couple} = 10^{-3}$,
showing that the final photon state is converged.

As the photon moves out, the polarization tensors (eqns.~[\ref{eq:dielectric_tensor}]-[\ref{eq:permeability_tensor}]) 
smoothly approach unity, with a consequent
freezing of the electric field vector. Figure~\ref{f:trajectory_error}a shows that the 
absolute value of the spatial components of $\mathbf{A}$ relative to the magnetic axis
cease to vary significantly at radii $\sim 200R_{\rm NS}$. Nonetheless, a quantitative criterion is needed to
establish freezing in practice. As a figure of merit, we measure the norm of the difference
between the polarization vector before and after each spatial step
\begin{equation}
\label{eq:A_error}
|\Delta \mathbf{A} | = | \mathbf{A}(\mathbf{r} + \Delta \ell \hat k) - \mathbf{A}(\mathbf{r}) |,
\end{equation}
and then proceed to multiply this quantity by the factor $(r/\Delta \ell)$, obtaining an
estimate of the change in $\mathbf{A}$ over a scale height in magnetic field strength. The result is
shown in Figure~\ref{f:trajectory_error}b. 
The dashed line shows the point where $|\Delta \mathbf{A}| (r / \Delta \ell) = 10^{-3}$, 
which we establish as our fiducial criterion for polarization freezing.
As shown by Figure~\ref{f:error_trajectory_k}, this rather conservative limit is located 
at distances $<1000R_{\rm NS}$ from the star for most trajectories, safely below the light cylinder
by at least an order of magnitude. It also lies below the point where the plasma
contribution would become comparable to that of the magnetized vacuum (eq.~[\ref{eq:dielectric_ratio}]).

Figure~\ref{f:trajectory_error}(c)-(d) also illustrates how normal modes become coupled in
a non-radial trajectory. Panel (c) shows the normalized intensity in the O-mode,
$|A_O|^2 = |\mathbf A\cdot \hat e_{O}|^2$ with $\hat e_{O}$ the O-mode eigenvector,
while panel (d) shows the normalized Stokes V component (eq.~[\ref{eq:stokes_V}]). 
Modes begin to couple within a factor 2 of the polarization limiting radius (eq.~[\ref{eq:r_pl}]),
before the electric field components freeze. This particular trajectory asymptotes to a $\sim 10\%$
amplitude in the orthogonal mode, and a circular polarization component of the same magnitude,
showing the importance of solving equations~(\ref{eq:prop_ax_general})-(\ref{eq:prop_ay_general}).

As a further test of our implementation, we have calculated the polarization signal as a function
of phase for two configurations with $\Delta \phi_{\rm NS} = 0$ (no scattering), light bending, and 
other parameters mirroring Figures~7 and 8 of \citet{heyl2003}\footnote{The panel in the lower left
could only be matched to the corresponding Figure in \citet{heyl2003} by changing the orientation angles.}. 
The latter are equivalent to
a configuration with stellar radius $R_{\rm NS} = 10$~km, a polar magnetic field $B_{\rm pole} = 2\times 10^{12}$~G, 
and monoenergetic photons with energy $\hbar \omega = 6.626$~eV and $0.6626$~keV, respectively.
\begin{figure}
\includegraphics*[width=\columnwidth]{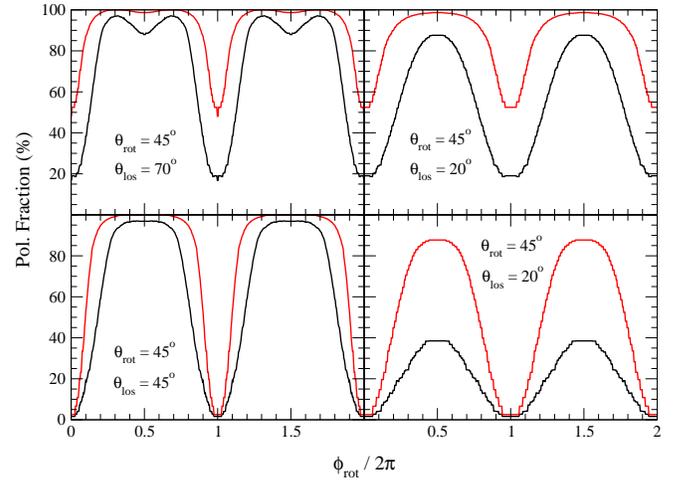}
\label{f:heyl_test}
\caption{Polarization fraction for monoenergetic photons and dipolar field (no scattering) as a 
function of phase for different relative orientations of magnetic axis, rotation axis, 
and line-of-sight (eqns.~[\ref{eq:murot}] and [\ref{eq:mulos}]). Parameters are $R_{\rm NS} = 10$~km,
$B_{\rm pole} = 2\times 10^{12}$~G, and $\hbar \omega = 6.626$~eV (black) and $0.6626$~keV (red). 
Results agree with Figures~7 and 8 of \citet{heyl2003}.}
\end{figure}

\subsection{Postprocessing}

\begin{deluxetable*}{lcccllc}
\tablecaption{Model Parameters\label{t:models}}
\tablewidth{0pt}
\tablehead{
\colhead{Name\tablenotemark{a}} &
\colhead{Dist.\tablenotemark{b}} &
\colhead{$\Delta \phi_{\rm NS}$ (rad)} &
\colhead{$\gamma_{\rm max}$} &
\colhead{Seed Pol.} &
\colhead{$\theta_{\rm cap}$ ($^o$)} &
\colhead{Fig.}
}
\startdata
t10\,g20\,c90\,Eu   & uni & 1   & 2           & E-mode & 90 & 7-20\\
\noalign{\smallskip}
t10\,g20\,c90\,Eb   & bid & 1   & 2           & E-mode & 90  & 12 \\
\noalign{\smallskip}
t08\,g24\,c90\,Eu   & uni & 0.8   & 2.4       & E-mode & 90 & 13,18\\
t12\,g18\,c90\,Eu   & uni & 1.2   & 1.8       & E-mode & 90 & 13,18\\
t14\,g17\,c90\,Eu   & uni & 1.4   & 1.65      & E-mode & 90 & 13,18\\
t16\,g16\,c90\,Eu   & uni & 1.6   & 1.6       & E-mode & 90 & 13,18\\
\noalign{\smallskip}
t10\,g20\,c90\,Ou   & uni & 1   & 2           & O-mode & 90 & 15,19\\
\noalign{\smallskip}
t10\,g20\,c5b\,Eu   & uni & 1   & 2           & E-mode & 5 (both)   & 16,20\\
t10\,g20\,c5s\,Eu   & uni & 1   & 2           & E-mode & 5 (south)  & 16,20\\
t10\,g20\,c5p\,Eu   & uni & 1   & 2           & E-mode & 5 (planar) & 16,20
\enddata
\tablenotetext{a}{We do not list here models with $\Delta \phi=0$, or test calculations.}
\tablenotetext{b}{Directionality of the particle energy distribution (eq.~[\ref{eq:pl_distribution}]), with
\emph{uni} denoting unidirectional, and \emph{bid} bidirectional, respectively.}
\end{deluxetable*}

Once the photon polarization is frozen, we store its frequency, position, direction, and polarization
vectors. In contrast to \citet{fernandez2007}, we do not convolve a frequency shift distribution
with a blackbody function, but instead directly draw photons from a Planck distribution. The convolution
of a frequency independent shift distribution with a source function is
incompatible with the fact that the polarization evolution equations depend explicitly on
the photon frequency through $k_0$ (eqns.~[\ref{eq:prop_ax_general}]-[\ref{eq:prop_ay_general}]).

In addition to the aforementioned variables, we keep track of the number of scatterings a photon 
has had before escaping, and construct separate histograms labeled by this parameter.

The standard parameterization of polarization degrees of freedom is a set of four Stokes parameters
(e.g., \citealt{rybicki2004}). Choosing the photon direction as the $\hat z$ axis, there is
an additional degree of freedom in their definition: the rotation angle around this axis. Choosing a 
coordinate system where the magnetic field is initially in the $x-z$ plane, we can write
\begin{eqnarray}
I & = & |A_x|^2 + |A_y|^2\\
Q & = & |A_x|^2 - |A_y|^2\\
U & = & 2{\rm Re}\left(A_x\,A_y^*\right)\\
\label{eq:stokes_V}
V & = & 2{\rm Im}\left(A_x\,A_y^*\right),
\end{eqnarray}
where the star stands for complex conjugation, and $I$ measures the intensity, $Q$ and $U$
linear polarization, and $V$ circular polarization.
Standard observables are
the linear polarization fraction\footnote{We focus on linear over full polarization 
fraction because photoelectric polarimeters will
not be sensitive to circular polarization (e.g., \citealt{swank2010}).}
\begin{equation}
\label{eq:pol_fraction}
\Pi_L = \frac{\sqrt{Q^2 + U^2}}{I},
\end{equation}
and the polarization angle
\begin{equation}
\label{eq:pol_angle_def}
\chi = \frac{1}{2}\arctan\left(\frac{U}{Q} \right).
\end{equation}
Rotation of the Stokes parameters around the $z$-axis is straightforward (e.g., \citealt{chandrasekhar1960}),
and involves only $Q$ and $U$. For phase dependent quantities, we perform a rotation such that
the $x$-axis is in the plane of the line of sight and the rotation axis.

Quantities as a function of phase are obtained by performing suitable geometric transformations
(Appendix~\ref{s:rotations}), which have as input parameters the two orientation angles
\begin{eqnarray}
\label{eq:murot}
\cos\theta_{\rm rot} & = & \hat M \cdot \hat \Omega\\
\label{eq:mulos}
\cos\theta_{\rm los} & = & \hat O \cdot \hat \Omega,
\end{eqnarray}
where $\hat M$, $\hat \Omega$, and $\hat O$ are the magnetic axis, rotation axis, and 
line of sight direction, respectively\footnote{These two angles are denoted respectively by 
$\xi$ and $\chi$ in \citet{nobili2008a}.}.

\subsection{Model Parameters}

Table~\ref{t:models} shows the sequence of models investigated in this paper. All of them cover the energy
range $0.04-40$~keV. We run two versions of each. One with 40 bins per decade in energy and 64 bins in
$\cos\theta_{\rm k}$, optimized for angle-averaged studies as a function of energy. We also compute
models with 10 bins per decade in energy and 256 bins in angle, to compute frequency integrated
lightcurves. Because the magnetic field
is axisymmetric, we average our results in $\phi_k$. The number of
seed photons is such that there are $10^4$ per energy-angle bin. 
The stellar radius is assumed to be $R_{\rm NS} = 10$~km, the Schwarzschild radius
$R_{\rm NS}/r_s = 3$, the magnetic field at the poles $B_{\rm pole} = 10^{14}$~G, 
and the temperature of the seed blackbody measured at infinity 
$kT_\infty = (1-r_s/R_{\rm NS})^{1/2}\, kT = 0.4$~keV.
Other parameter combinations for each model are listed in Table~\ref{t:models}. Model names refer to their twist 
angle (t), maximum Lorentz factor (g), polar cap radius (c),
seed polarization (E/O), and directionality (u/b).

\begin{figure*}
\includegraphics*[width=\textwidth]{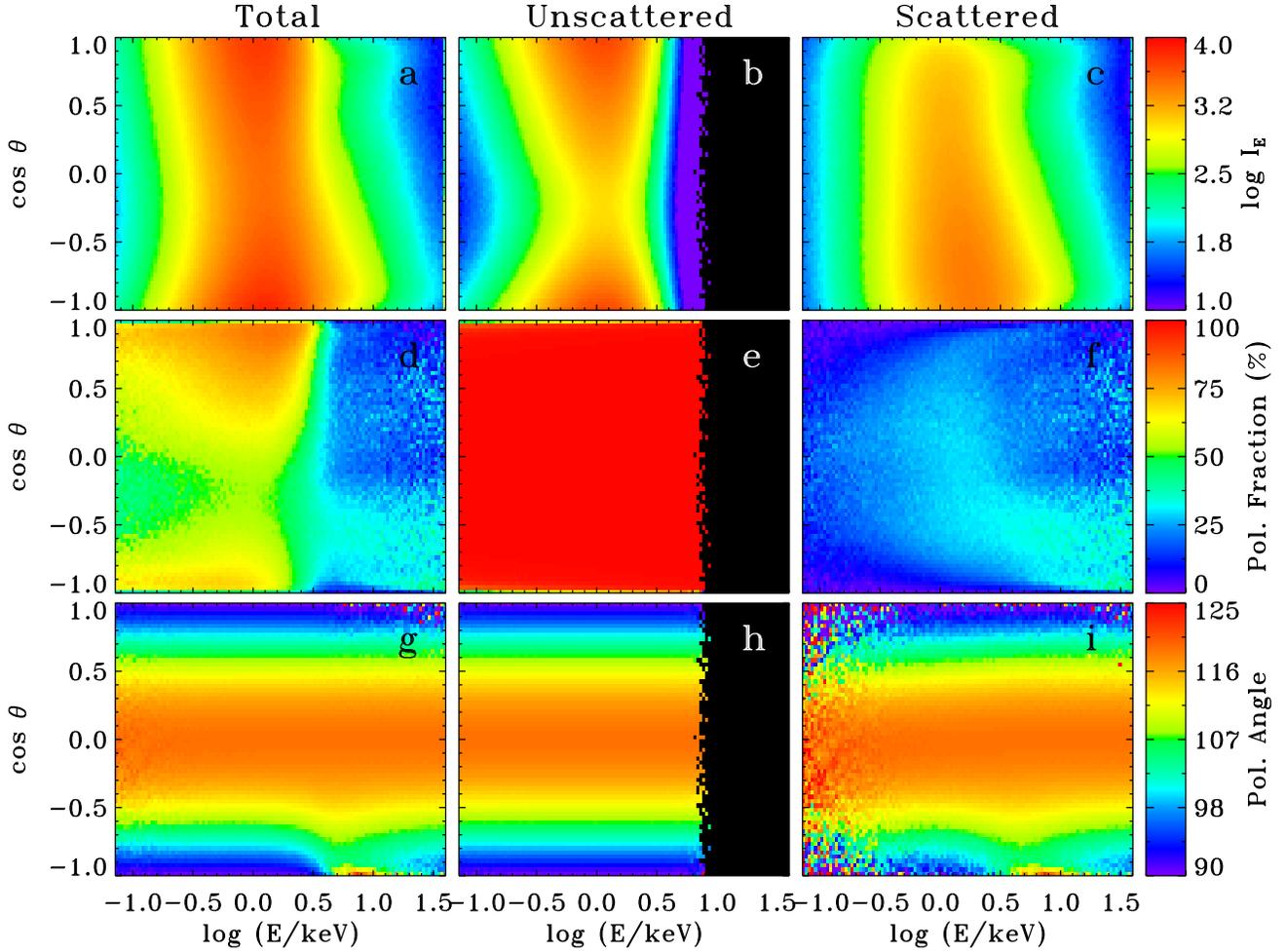}
\caption{
Observables measured at infinity for model t10\,g20\,c90\,Eu, as
a function of photon energy and magnetic colatitude. Each panel shows,
from top to bottom, number of photons collected, linear polarization
fraction (eq.~[\ref{eq:pol_fraction}]), and polarization angle (eq.~[\ref{eq:pol_angle_def}]), 
respectively. Left, middle, and
right panels correspond to the total, unscattered, and scattered
contributions, respectively.  Regions of parameter space with no
counts are shown in black.}
\label{f:2d_histogram}
\end{figure*}

\section{Results}                                                                                   
\label{s:results}

Since one of our primary goals is to provide predictions for future
X-ray polarimeters  like GEMS, we focus on the behavior of 
the linear polarization fraction (eq.~[\ref{eq:pol_fraction}]) and 
polarization angle (eq.~[\ref{eq:pol_angle_def}]) in relation to the total intensity, as different
system parameters are changed.
Phase-integrated and phase-dependent observables
provide complementary information about the system, and we consider
each class of measurement in turn.  
Due to rotation, both sets of observables depend on the 
relative orientation of rotation axis, magnetic axis, and line-of-sight, as 
parameterized by the angles $\theta_{\rm rot}$ and $\theta_{\rm los}$ 
(eqns.~[\ref{eq:murot}] and [\ref{eq:mulos}], respectively).  
Therefore, we begin by discussing the emission in the corotating frame, where the physics giving rise to the
polarization signal is most easily understood.

\subsection{Polarized Emission in the Corotating Frame}
\label{s:corotating}

We first focus on the basics of the angular and frequency distribution
of the polarized emission in the corotating frame.  
Figure \ref{f:2d_histogram} shows two-dimensional histograms of
intensity, polarization fraction, and polarization angle measured at
infinity, as a function of photon energy and magnetic colatitude, for
model t10\,g20\,c90\,Eu. For illustration, we have separated out in
different columns the contributions from scattered and unscattered
photons, plus their sum.

Panels (a)-(c) 
of Figure \ref{f:2d_histogram} 
show the number of photons collected at infinity. The
unscattered distribution has 
an energy dependence close to that of a blackbody\footnote{The optical depth (eq.~[\ref{eq:dtau}]) depends
implicitly on frequency through the resonant momenta $(\gamma\beta)^{\pm}$.} and
an angular dependence given by $I_{\rm unscatt} = I_0
e^{-\tau(\theta)}$, where $I_0$ is the surface intensity and $\tau
\sim \sin^2\theta \Delta \phi/\bar{\beta}$ the total optical depth
(the integral along the line-of-sight of eq.~[\ref{eq:dtau}]). The
exact dependence of the latter on magnetic colatitude and the
unidirectional particle velocity distribution introduce a slight north-south
asymmetry.  The scattered component, on the other hand, peaks at the
southern hemisphere. This results from
the fact that photons with a large positive energy shift move out of the
magnetosphere
when they are generated close to the magnetic equator, where the
magnetic field points downwards. Upscattered photons moving northward
are re-entering the magnetosphere (see below).
At fixed magnetic colatitude, the frequency dependence
is smooth curve and extends to both sides of the 
original
blackbody peak.

Panels (d)-(f) of Figure \ref{f:2d_histogram} show the corresponding polarization fractions. The
unscattered component reflects the polarization distribution of seed
photons, in this case 100\% E-mode. The scattered component has a
characteristic polarization fraction of $\sim 30\%$, which arises
mostly from mode exchange (see below). The total polarization fraction
peaks at the north pole, because more unscattered photons leave in
that direction, and they are more polarized. This is the consequence
of having a uni-directional particle energy distribution, which causes
a break of the north-south symmetry. Thus, the polarization fraction
carries a signature in both energy and angle.

From equation (\ref{eq:mode_switch_prob}), one can see
that photons are
more likely to be in the E-mode state immediately after scattering, for all
$|\mu_r| < 1$.  This, coupled with nearly adiabatic propagation, means that
E-mode polarization dominates the scattered emission for all values of
the energy and magnetic colatitude. If photons were to scatter only
once, the net polarization fraction would be
\begin{equation}
\label{eq:linearpol_singlescatt}
\Pi_L = \int\,\left[\frac{1-(\mu_r^\prime)^2}{1 + (\mu_r^\prime)^2}\right] [1+(\mu_r^\prime)^2]\,\totd \mu_r^\prime
\bigg/ \int\,[1+(\mu_r^\prime)^2]\,\totd\mu_r^\prime = 0.5,
\label{eq:rest_frame_pol}
\end{equation}
where the weight $[1+(\mu_r^\prime)^2]$ comes from the total differential cross-section (eq.~[\ref{eq:diff_cross_sec}]).
The characteristic polarization fraction of $\sim 30\%$ observed at infinity arises
from multiple scattering. In particular, E-mode photons have a larger cross-section than O-mode 
photons at all angles because of the overlap $|e_\pm|^2$ 
(eqns.~[\ref{eq:overlaps_outgoing}] and [\ref{eq:non_adiabatic_overlap}]), and hence they are more likely to
undergo further scatterings, decreasing the overall polarization fraction. 

\begin{figure}
\includegraphics*[width=\columnwidth]{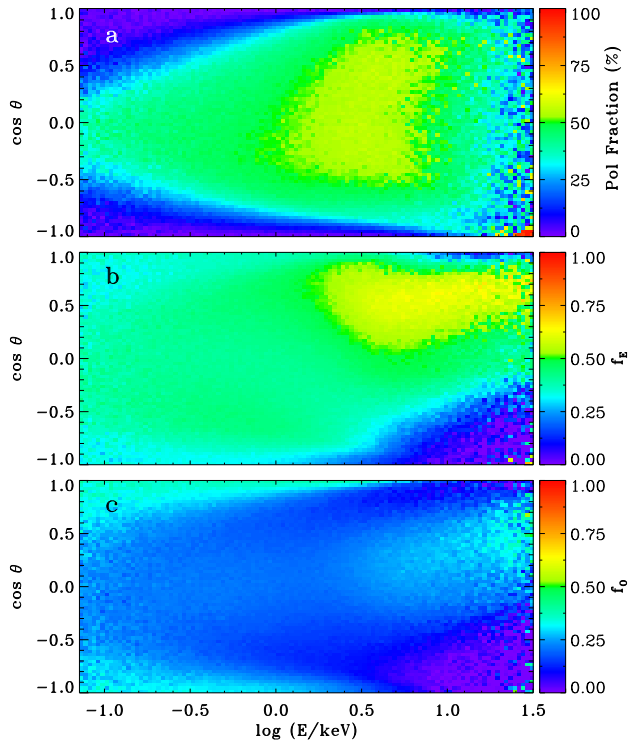} 
\caption{
\emph{Top:} Polarization fraction in the first scattering order for a run with the same
parameters as model t10\,g20\,c90\,Eu, but forcing photons to escape after one
scattering. \emph{Middle and Bottom:} Fraction of re-scattered
E- and O-mode photons, respectively, obtained by subtracting the normal
mode intensities of model t10\,g20\,c90\,Eu (which undergoes multiple scattering) 
from the corresponding values from the run shown in the top panel (with a single scattering).
As E-mode photons have a larger cross-section, they undergo relatively more re-scatterings than
O-mode photons and hence the polarization fraction decreases from the upper limit given by
equation~(\ref{eq:linearpol_singlescatt}).
}
\label{f:firstord}
\end{figure}

To show this effect, Figure~\ref{f:firstord}(a) displays the polarization fraction of the scattered
component in a run with the same parameters
as model t10\,g20\,c90\,Eu (Figure~\ref{f:2d_histogram}) but this time forcing photons to escape after
their first scattering. The average polarization fraction
is consistent with what equation~(\ref{eq:rest_frame_pol}) predicts. One can take the difference between the normal
mode intensities in this run and those from the first scattering order in 
model t10\,g20\,c90\,Eu to find the fraction of re-scattered
photons. The resulting numbers for E- and O-mode are shown in panels (b) and (c) of Figure~\ref{f:firstord}.
As expected, E-mode photons undergo relatively more re-scatterings than O-mode photons, decreasing the
overall polarization fraction of the first scattering order while still dominating the output. 
The bulk of re-scattering affects photons that move towards the north magnetic pole and have a high
energy. The maximal frequency boost imparted by equation~(\ref{eq:omega_doppler}) is 
$(1+\beta\mu_{\rm in})/(1-\beta\mu_{\rm out})$,
with $\mu_{\rm in}>0$ and $\mu_{\rm out}>0$ the ingoing and outgoing cosine between photon direction and
magnetic field ($\beta > 0$ in the uni-directional case). Hence these photons correspond to those
who first scattered closer to the south magnetic axis, where the magnetic field points mostly northward,
and whose value of $\mu$ changed from negative to positive. After their first scattering, these photons
move inwards, being most likely to scatter again. An
analogous argument shows why high-energy photons
going southward after scattering are the most likely to escape. 

Panels (g)-(i) of Figure~\ref{f:2d_histogram} show the polarization angle. 
It is immediately evident that
this quantity is mostly independent of energy for this particular
case, varying only with magnetic colatitude, aside from a small region of scattered
emission near the south pole.  Absent twist, the magnetic field is
dipolar and parallel to the magnetic axis along the line-of-sight
This would yield a polarization angle of $90^\circ$ for the
E-mode and $0^\circ$ for O-mode.  A positive $\Delta \phi_{\rm NS}$
introduces a dependence on magnetic colatitude, increasing these values
away from the poles. The functional form is equal, within statistical errors, 
to the angle that the magnetic field makes with the magnetic axis.
For model t10\,g20\,c90\,Eu and measured relative to the magnetic axis,
the polarization angle is
$\chi = \arctan{(B_\phi/B_\theta)} + \pi/2$.

The weak dependence on energy results from E-mode polarization
dominating the scattered component at most angles and energies,
combined with the assumption of E-mode seed photons in the
t10\,g20\,c90\,Eu model.  
This dominance of the E-mode in the scattered
emission persists even if $100\%$ O-mode seed photons are used.
The only difference is a $90^\circ$ jump as a function of energy when transitioning 
to the unscattered component. The polarization vector follows the local magnetic field
direction adiabatically until the coupling region at $r\sim r_{\rm pl}$ (\S\ref{s:transfer}). 
Hence, the polarization angle of the scattered component at most latitudes will trace the
magnetic field direction at $r\sim r_{\rm pl}\sim 100R_{\rm NS}$ (eq.~[\ref{eq:r_pl}]). 

\begin{figure}
\includegraphics*[width=\columnwidth]{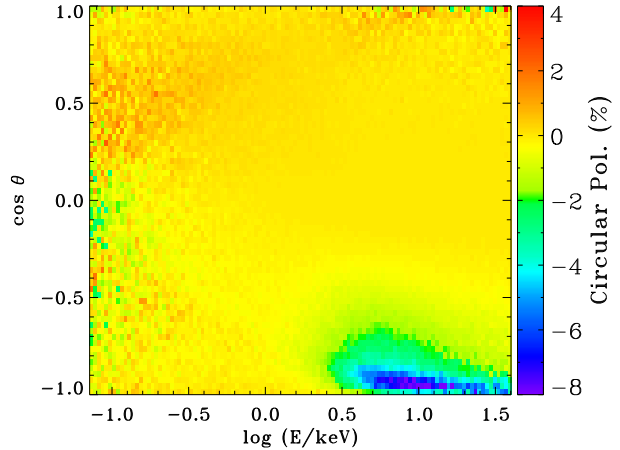}
\caption{
Circular polarization fraction measured at infinity (normalized Stokes V parameter, eq.~[\ref{eq:stokes_V}]) 
for model t10\,g20\,c90\,Eu, as a function of photon energy and magnetic colatitude.
}
\label{f:vcirc}
\end{figure}

Figure~\ref{f:vcirc} shows the degree of circular polarization for model
t10\,g20\,c90\,Eu as a function of energy and magnetic colatitude. 
It is produced almost exclusively by the scattered component, and
its value deviates from zero by a few percent throughout the energy-colatitude
plane, except in regions close to the magnetic poles at high energies.
The maximum absolute value of about 8\% is reached near the south pole, towards
which most scattered photons with a positive energy shift escape.

In principle, any photon moving non-radially will develop a circular polarization
component as it traverses the magnetosphere (e.g., Figure~\ref{f:trajectory_error}[c]). To see this,
one can write an equation for the rate of change of $V$ along the photon 
trajectory, by combining equations~(\ref{eq:prop_ax_general}), (\ref{eq:prop_ay_general}), 
and  (\ref{eq:stokes_V}). After some algebra, one finds that to leading order in 
$\delta = \alpha_{\rm em}(B/B_{\rm QED})^2/(45\pi)$ one has
\begin{equation}
\frac{\partial V}{\partial z} = (q+m)\sin^2\theta_{kB}\cos 2\phi_{kB}\,U + O(\delta^2),
\end{equation}
where $\theta_{kB}$ and $\phi_{kB}$ are respectively the polar and azimuthal angle of the magnetic field in a 
coordinate system where $\hat k = \hat z$ and $\hat B$ is in the x-z plane. 
If the polarization vector is in a normal mode state,
then $U=0$ and no circular component is generated. Similarly, $\totd V/\totd z = 0$
if photons propagate parallel to the magnetic field. 

\begin{figure}
\includegraphics*[width=\columnwidth]{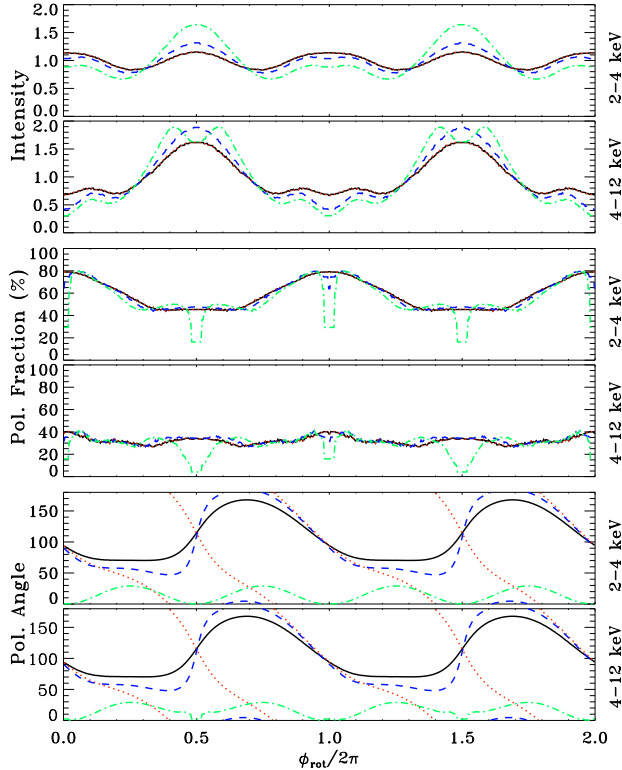} 
\caption{
Intensity (top), polarization fraction (middle), and polarization
angle (bottom) as a function of 
rotational phase $\phi_{\rm rot}$
in two energy bands (2-4 keV and
4-12 keV) for
model
t10\,g20\,c90\,Eu
and various viewing
geometries.  The curves correspond to pairs of angles ($\theta_{\rm
  rot}$, $\theta_{\rm los}$) of ($45^\circ$, $70^\circ$: solid/black),
($70^\circ$, $45^\circ$: dotted/red), ($60^\circ$, $70^\circ$:
dashed/blue), and ($90^\circ$, $90^\circ$: dot-dashed/green).
For illustration, two full periods are shown.}
\label{f:phase_ang}
\end{figure}

The fact that the unscattered component has $V$ almost identically zero everywhere 
is due to the fact that a given non-radial trajectory originating from the star always
has a counterpart for which the magnetic 
field rotates in the other direction along the trajectory,
cancelling out the net contribution to $V$ if they have the same intensity.
The scattered component, however, is not isotropic and thus such cancellation does not occur, 
leaving a residual circular polarization signal.
If plasma polarization were to become dominant at some radius comparable to $r_{\rm pl}$ 
(eq.~[\ref{eq:r_pl}]), then the circular polarization fraction could be much higher and 
substantially modify the outgoing linear polarization signal.
\vspace{0.5in}

\subsection{Phase-resolved Emission}
\label{s:phase_resolved}

Ideally, one would like to observe the system from all angles in the
corotating frame and generate observed maps as a function of magnetic
colatitude for comparison with plots like Figure
\ref{f:2d_histogram}. In practice, we have a fixed line of sight for
any single source, so that the viewing angle corresponds to a time
varying magnetic colatitude as a function of rotational phase
$\phi_{\rm rot}$. 
Due to this relatively direct
relationship with the corotating frame observables, we find it
pedagogically advantageous to discuss phase-resolved observables
first.  This relationship also means that polarization measurements as
a function of rotational phase yield the most detailed information
about the system.  Hence, they are the most useful observational
constraints for disentangling geometrical effects from the physics of
the 
surface and magnetosphere.  However, they may
necessitate much longer integration times to achieve useful
signal-to-noise.  We discuss prospects for obtaining such
phase-resolved measurements with GEMS in \S\ref{s:gems}.

Figure \ref{f:phase_ang} shows several light curves corresponding the
to the t10\,g20\,c90\,Eu model for different 
lines-of-sight
and magnetic axis
orientations. The phase is defined so that the north and south magnetic poles
cross the 
line-of-sight at $\phi_{\rm rot}=0$ and $\pi$, respectively.
Two full periods are shown for illustration.  The intensity,
polarization fraction, and polarization angle are computed after
integrating the emission over a soft energy band (2-4 keV) that is
dominated by direct emission, and a hard band (4-12 keV) which is
mostly scattered flux.
The normalization is chosen so that the average over phase is unity.

\begin{figure}
\includegraphics*[width=\columnwidth]{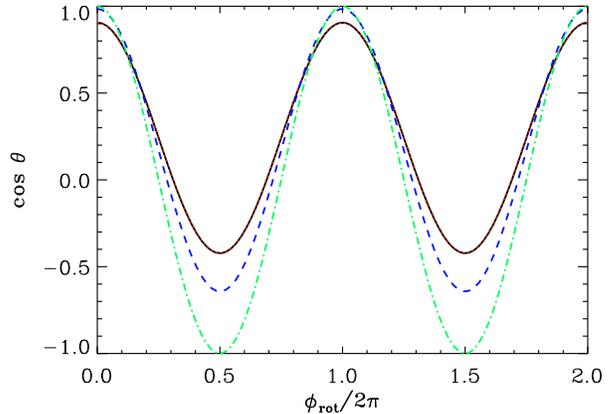}
\caption{
Magnetic colatitude 
swept by the line-of-sight
as a function of rotational phase for various
viewing geometries.  The color and line styles have the same meanings
as in Figure~\ref{f:phase_ang}.}
\label{f:phase_map}
\end{figure}

The variation of the total intensity and polarization fraction follows
directly from the two dimensional map in Figure~\ref{f:2d_histogram},
as there is a simple mapping between rotational phase and magnetic
colatitude (eq.~[\ref{eq:rotphase}]).  Figure~\ref{f:phase_map}
shows this mapping for the same set of orientations ($\theta_{\rm
  rot}$, $\theta_{\rm los}$) that are plotted in Figure~\ref{f:phase_ang}.  
Orientation effects depend mostly on
the maximum (most negative $\cos\theta$) observable colatitude, 
because the
polarization fraction tends to be a minimum and the total intensity a
maximum near the south magnetic pole.  The mapping is invariant under
the exchange $\theta_{\rm rot} \leftrightarrow \theta_{\rm los}$, so
the solid black ($\theta_{\rm rot}=45^\circ$, $\theta_{\rm
  los}=70^\circ$) and dotted red curves ($\theta_{\rm rot}=70^\circ$,
$\theta_{\rm los}=45^\circ$) are identical.

The orthogonal rotator viewed at $\theta_{\rm los}=90^\circ$ gives the
largest modulation.  In this case the observer sees one of the
magnetic poles directly at $\phi_{\rm rot}$ equal to integer multiples
of $\pi$.  The optical depth to scattering is lowest along the poles,
leading to dips in the total intensity at these phases.  The
polarization fraction is also low since the photon wave vector lies parallel
to the magnetic field and there is no preferred direction for the
polarization.

The gross, qualitative features of the total intensity and
polarization fraction phase curves are largely insensitive to the
orientation.  There is a strong peak near $\phi_{\rm rot}=\pi$ and
generally a weaker one near $\phi_{\rm rot}=0$, when the south and
north magnetic poles (respectively) cross the 
line-of-sight.  The contrast is particularly clear in the hard energy
band which is dominated by the scattered emission.  However, in the
soft band, which is dominated by the unscattered component the
polarization fraction peaks near $\phi_{\rm rot}=0$ and has a minimum
near $\phi_{\rm rot}=\pi$.

\begin{figure}
\includegraphics*[width=\columnwidth]{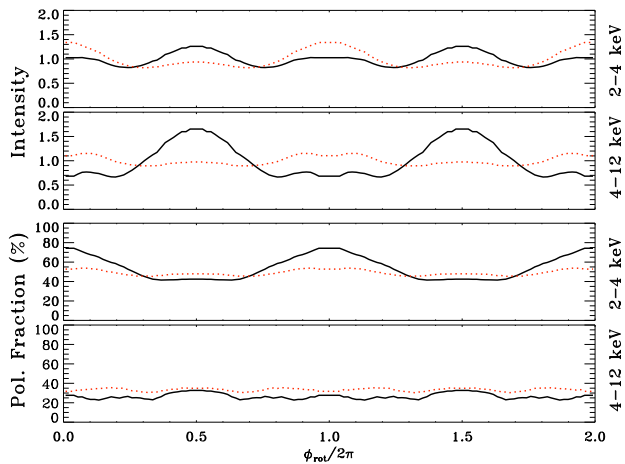}
\caption{
Intensity (top) and polarization fraction (bottom) as a function of
phase in two energy bands (2-4 keV and 4-12 keV). The curves
correspond to models t10\,g20\,c90\,Eu (black solid) and
t10\,g20\,c90\,Eb (red dotted). The former assumes that the particle distribution is
unidirectional, while that of the latter is bidirectional. Magnetic and viewing
orientations corresponding to $\theta_{\rm rot}=45^\circ$ and
$\theta_{\rm los}=70^\circ$ are assumed.}
\label{f:phase_distdir}
\end{figure}

The half period phase shift between the polarization fraction and the
total intensity results from the assumption of a unidirectional photon
distribution. Figure~\ref{f:phase_distdir} shows a comparison of the
t10\,g20\,c90\,Eu and t10\,g20\,c90\,Eb models.  These differ only in
their treatment of the particle distribution, with the latter having a
bidirectional charge distribution.  The largest contrast arises when
comparing the intensity in the hard band with the polarization
fraction in the soft band.  The intensity and polarization fraction of
the bidirectional model is completely symmetric in magnetic colatitude,
so the differences between $\phi_{\rm rot}=0$ and $\pi$ are purely
orientation effects in this case\footnote{The observable
colatitude range is $[25^\circ,115^\circ]$}.  In contrast to the unidirectional
model, the peaks in intensity and polarization fraction coincide.
Thus, one can distinguish whether the magnetosphere is filled with
electron-positron or electron-ion pairs because this difference in
particle distribution manifests itself as a break in the north-south
symmetry for the unidirectional model.

As mentioned in \S\ref{s:corotating}, the polarization angle 
is determined primarily by the projection of the magnetic field at $r\sim r_{\rm pl}$ 
onto the line-of-sight.  
In phase resolved plots, we have chosen the zero point for this angle to 
be the rotation axis, so the absolute value will differ from that shown
in Figure~\ref{f:2d_histogram} by a geometric transformation involving
$\theta_{\rm rot}$, $\theta_{\rm los}$, and the rotational phase $\phi_{\rm rot}$
(eqns.~[\ref{eq:rotphase}]-[\ref{eq:stokesang}]). This means that the
phase curve of polarization angle is not affected by the invariance to
$\theta_{\rm rot}\leftrightarrow \theta_{\rm los}$ exchange of the
phase-colatitude mapping, as there is an additional rotation of the polarization
vector as the magnetic axis rotates around the line-of-sight.
This is seen
most clearly by comparing the $(\theta_{\rm rot}=45^\circ,
\theta_{\rm los}=70^\circ)$ and $(\theta_{\rm rot}=70^\circ,
\theta_{\rm los}=45^\circ)$ curves in Figure~\ref{f:phase_ang}.  
Although the intensity and polarization fraction phase curves are identical, the variation with
polarization angle differs greatly between the two orientations.  
In general, for $\theta_{\rm rot} < \theta_{\rm los}$, the polarization
oscillates about a mean value, while for $\theta_{\rm rot} > \theta_{\rm
  los}$ it sweeps through the full allowed range of $180^\circ$.
This has important implications for the phase-averaged polarization,
which we discuss below.

\begin{figure}
\includegraphics*[width=\columnwidth]{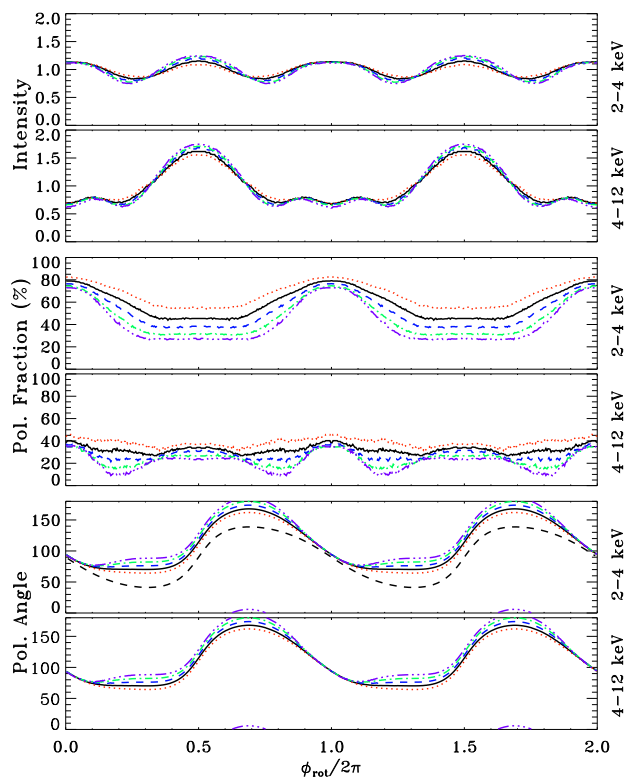}  
\caption{
Intensity (top), polarization fraction (middle), and polarization
angle (bottom) as a function of phase in two energy bands (2-4 keV and
4-12 keV)
for different twist angles.
The curves correspond to models
t10\,g20\,c90\,Eu (solid, black), t08\,g20\,c90\,Eu (dotted, red),
t12\,g20\,c90\,Eu (blue, dashed) t14\,g20\,c90\,Eu (green, dot-dashed), and
t16\,g20\,c90\,Eu (violet, triple-dot-dashed).  Magnetic and viewing
orientations corresponding to $\theta_{\rm rot}=45^\circ$ and
$\theta_{\rm los}=70^\circ$ are assumed. For comparison, the long-dashed black
line shows the polarization angle as a function of phase for a run with no
twist and thus a purely dipolar field. }
\label{f:phase_twist}
\end{figure}

Figure \ref{f:phase_twist} shows the variation of the phase-resolved
observables as twist angle $\Delta \phi_{\rm NS}$ and maximum Lorentz factor
$\gamma_{\rm  max}$ of the momentum distribution (eq.~[\ref{eq:pl_distribution}]) 
are varied. We are primarily interested in the variation with
$\Delta \phi_{\rm NS}$, but we vary $\gamma_{\rm max}$ simultaneously
(see Table~\ref{t:models}) to keep the power law slope of the spectrum
as uniform as possible.
As twist angle increases, the intensity increases for all energies above
$\sim 2$~keV (c.f.~\S\ref{s:phase_average}) as the fraction of scattered
photons increases. Thus the peak in the intensity
near $\phi_{\rm rot}=\pi$ becomes stronger in both bands.  

The polarization fraction shows the opposite trend with twist angle,
however.
As $\Delta \phi_{\rm NS}$ increases, the polarization fraction gets
smaller at every phase in both the soft and hard bands.  For the soft
band, the primary explanation is that scattered photons have lower
polarization than unscattered, so that a larger fraction of scattered
photons yields a lower overall polarization fraction.  As with the
intensity variation, the effect is again largest at $\phi_{\rm rot}=\pi$, 
when the south pole crosses the line-of-sight.  This
result is qualitatively similar to that found by \citet{nobili2008a}
using adiabatic propagation and taking the relative difference between
the normal mode intensities to obtain polarization
fractions. \citet{nobili2008a} averaged over magnetic colatitude and
energy to obtain their result, and offered a similar interpretation.

However, it is notable that the drop in polarization fraction occurs
even in the hard band.  Unlike the soft band, it contains very little
highly-polarized, unscattered emission, so the above explanation does
not apply.  Instead, the drop in polarization with increasing twist
angle is due to the increased number of scatterings per photon, which remove
relatively more E-mode than O-mode photons from the escaping beam along a given
direction (see also \S\ref{s:corotating}). 

For the orientation of Figure~\ref{f:phase_twist},
the polarization angle oscillates about a mean
value near $130^\circ$, with an overall shape that depends weakly on
twist angle
but which becomes more asymmetric as this parameter is increased.
For reference, the black dashed curves in the panel corresponding
to the soft band shows the variation in the polarization angle with
phase for a model with $\Delta \phi_{\rm NS}=0$.  In this case the
oscillation is symmetric about $90^\circ$ because the field is
dipolar.
Hence, asymmetry in the polarization angle is
a direct indication of 
the presence of a twist angle at $r\sim r_{\rm pl}$,
and the magnitude of the asymmetry constrains the value of $\Delta \phi_{\rm NS}$.

\begin{figure}
\includegraphics*[width=\columnwidth]{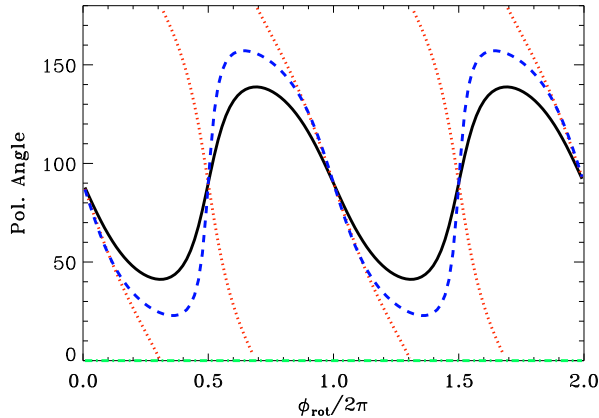}
\caption{Polarization angle as a function of phase for a model with no twist (no scattering)
and 100\% E-mode seed photons for the same orientations as in Figure~\ref{f:phase_ang}. 
Results for the soft and hard bands are identical due to lack of scattering, hence only one
panel is shown.
Note that the curves are symmetric every $1/2$ phase due to the dipolar geometry of the
zero-twist field. A net twist angle of the field lines introduces an asymmetry for most orientations.}
\label{f:polangle_notwist}
\end{figure}

To illustrate this effect in more detail, Figure~\ref{f:polangle_notwist} shows the polarization angle 
that is obtained when the twist angle is set to zero, for the same set of orientations
shown in Figure~\ref{f:phase_ang} (only one panel is shown, as the lack of scattering makes the
results identical in both bands).
Except for the orthogonal rotator, all curves in Figure~\ref{f:phase_ang} break the symmetry
at $1/2$ phase relative to Figure~\ref{f:polangle_notwist}, showing how the presence of
a net twist angle in the magnetosphere becomes directly  measurable through the polarization angle.
We emphasize that this effect is a more general consequence
of the properties of QED eigenmodes \citep{heyl2000}, and only indirectly due to
magnetospheric scattering in that a definite polarization mode becomes dominant.
It would be worth exploring under what circumstances it is possible to uniquely
reconstruct the magnetic field geometry from a given polarization angle curve;
this work however lies beyond the scope of this paper.

\begin{figure}
\includegraphics*[width=\columnwidth]{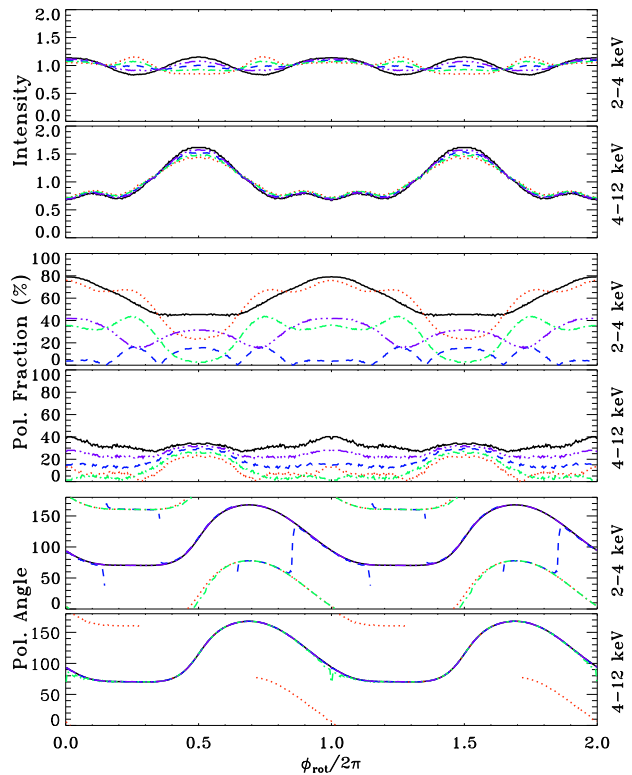}  
\caption{
Intensity (top), polarization fraction (middle), and polarization
angle (bottom) as a function of phase in two energy bands (2-4 keV and
4-12 keV) as the seed photon polarization is varied. The curves correspond
to models with seed distributions that are 100\% E-mode (solid,
black); 100\% O-mode (dashed, red); 50\% E-mode
+ 50\% O-mode (blue,
dashed); 25\% E-mode 
+ 75\% O-mode (green, dot-dashed); and
75\% E-mode
+ 25\% O-mode (purple, triple-dot-dashed).
Magnetic and viewing orientations corresponding to $\theta_{\rm
  rot}=45^\circ$ and $\theta_{\rm los}=70^\circ$ are assumed.}
\label{f:phase_seed}
\end{figure}

We now consider the dependence 
of observables
on the seed photon polarization, which
so far has been 100\%
E-mode. 
Figure~\ref{f:phase_seed} shows the variation of the phase-resolved
quantities
as the polarization of the seed photon distribution
varies.  Only two models with 100\% E-mode and O-mode seed polarization 
are directly computed
(t10\,g20\,c90\,Eu and t10\,g20\,c90\,Ou, respectively). 
The linearity of the Stokes parameters allows us to
combine these runs to get any partial polarization. Here we consider
an unpolarized (50\% E-mode, 50\% O-mode) seed distribution and a two
partially polarized (75\% E-mode and 25\% O-mode, 25\% E-mode and 75\%
O-mode ) to illustrate the trends.  The soft band intensity changes
because the optical depth has a different colatitude variation in the
pure O-mode model (t10\,g20\,c90\,Ou), due to the angular dependence in
the overlap function $|e_\pm|^2$ (eqns.~[\ref{eq:overlaps_outgoing}] and [\ref{eq:non_adiabatic_overlap}]).  
The hard band has a slightly weaker
peak for the pure O-mode model due to the reduced scattering fraction.
The unpolarized and partially polarized models lie between the pure
polarization models by construction.

The polarization fraction shows a somewhat stronger dependence on the
seed photon distribution.  For the pure O-mode model, the soft band
polarization fraction is much weaker that the pure E-mode case at
$\phi_{\rm rot}=\pi$.  This is 
because the 
transition from dominance of
unscattered O-mode seeds to E-mode dominated scattered emission 
occurs within this band.
As expected,
the unpolarized and partially polarized models yield lower
polarization fractions in the soft band, where unscattered photons
dominate.  The unpolarized model has a non-zero polarization fraction
solely because of the non-negligible scattered emission
towards lower energies.
In the hard band, the scattered emission is E-mode dominated for all models, and
the polarization fraction generally increases as the E-mode fraction
of the seed photon increases.

In the soft band, which is dominated by unscattered photons, the
phase curves of polarization angle for the models with pure E-mode and
pure O-mode seeds differ by exactly $90^\circ$.  The partially
polarized models follow the  curve of the dominant polarization, while
the unpolarized model switches between the two. In the hard band,
the pure E-mode and pure O-mode models yield identical polarization
angles at some phases and differ by $90^\circ$ at others.  These
discontinuous jumps 
occur in the 100\% O-mode seeds case due to a colatitude-dependent
transition from O-mode seeds to E-mode dominated scattered photons.
The partially polarized and unpolarized models follow the E-mode curve.

\begin{figure}
\includegraphics*[width=\columnwidth]{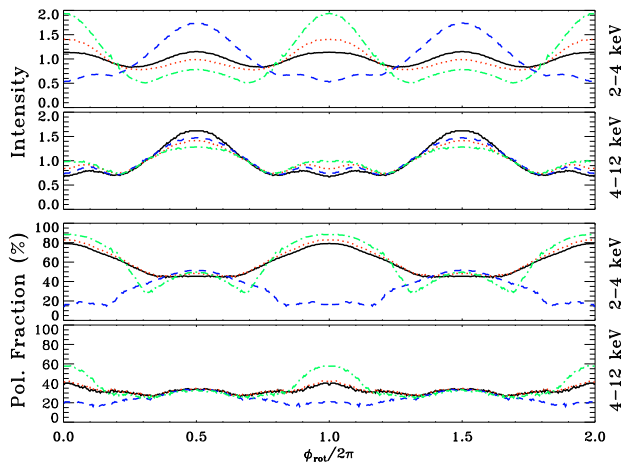}
\caption{
Intensity (top) and polarization fraction (bottom) as a function of
phase in two energy bands (2-4 keV and 4-12 keV) for various
assumptions about NS surface emission
and photon propagation.
The curves correspond to full surface emission with light bending
included (t10\,g20\,c90\,Eu, solid black), $5^\circ$ polar cap
on both magnetic poles with light bending (t10\,g20\,c5b\,Eu, dotted red),
$5^\circ$ polar cap with light bending but from the south pole only 
(t10\,g20\,c5s\,Eu, dashed blue), and $5^\circ$ polar caps from 
both poles and planar propagation (t10\,g20\,c5p\,Eu, dot-dashed green).
Magnetic and viewing orientations
corresponding to $\theta_{\rm rot}=45^\circ$ and $\theta_{\rm
  los}=70^\circ$ are assumed.}
\label{f:phase_cap}
\end{figure}

Thus far all the models have assumed uniform emission of seed photons
from the NS surface.  Figure \ref{f:phase_cap} illustrates the
effects of confining emission to polar caps of $5^\circ$ radius around
the magnetic poles.
In this case we only show the intensity and polarization fraction, as 
the polarization angle 
curves are identical.  The solid (black) curves correspond to emission
from the full surface (t10\,g20\,c90\,Eu) while the 
red-dotted curves refer to a model with polar caps around both magnetic
poles (t10\,g20\,c5bu). For the latter,
there are modest enhancements and reductions of the intensity at 
$\phi_{\rm rot}=0$ and $\pi$, respectively.  This results primarily from the
magnetic colatitude dependence of the scattering cross section, which
is lowest near the poles. 

Although there are these modest differences, the phase curves are
remarkably similar for the two models, given the large differences in
the fraction of the star that emits.  Even though the polar cap
emission is confined to a very small solid angle, 
light bending causes the NS to appear as a more isotropic emitter for
observers several stellar radii from the star where much of the
scattering occurs.  For comparison, the green (dot-dashed)
curves show a model with planar propagation and no light bending (t10\,g20\,c5pu).
The differences with the isotropic emitter are much larger and 
there is much clearer signature in the polarization fraction.

Finally, we consider a case in which only the south magnetic pole
radiates, including light-bending (t10\,g20\,c5su).  
This corresponds to the blue (dashed) curve in Figure~\ref{f:phase_cap}.  
Since we have chosen an orientation in which the
observer primarily sees the northern hemisphere, the lightcurve peaks
at $\phi_{\rm rot}=\pi$, when the south pole crosses the
line-of-sight.  When the north pole crosses our line-of-sight at
$\phi_{\rm rot}=0$, there are very few unscattered seed photons,
resulting in a low polarization fraction.

\subsection{Phase-averaged Emission}
\label{s:phase_average}

Averaging quantities the over stellar rotation
yields higher signal-to-noise.
Hence, phase-averaged
measurements are likely to be the first (or perhaps only) constraints
obtained when polarimeters become operational.  

\begin{figure}
\includegraphics*[width=\columnwidth]{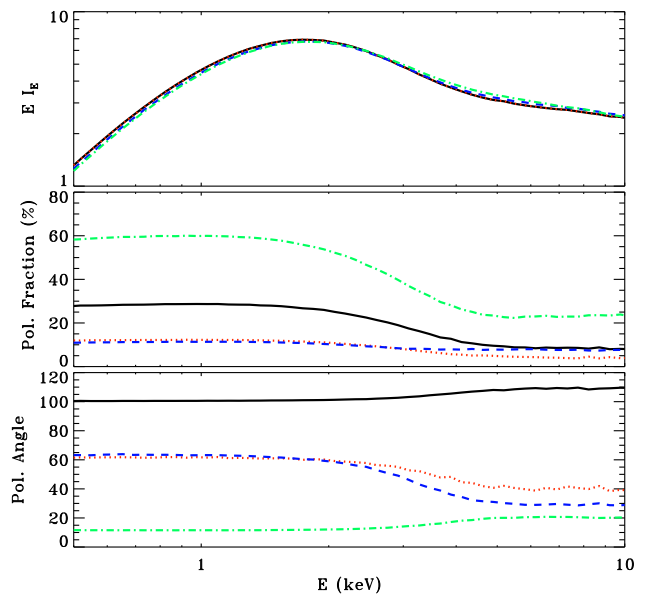}
\caption{
Phase-averaged intensity (top), polarization fraction (middle), and
polarization angle (bottom) as a function of photon energy for
model t10\,g20\,c90\,Eu
and various viewing geometries. The curves correspond to the same
orientation angles 
as in Figure~\ref{f:phase_ang}.}
\label{f:en_ang}
\end{figure}

Figure~\ref{f:en_ang} shows the t10\,g20\,c90\,Eu model for the same
orientation angles
as in Figure~\ref{f:phase_ang}.
Independent of 
geometric arrangement,
spectra show a decrease in the
polarization fraction with energy as they transition from
unscattered to scattering-dominated.
This transition occurs over the range $2-4$~keV, and involves
a decrease in the net polarization fraction.

Although 
this decrease
is generic, the precise
value of the polarization fraction and angle are orientation
dependent.  A comparison of the middle panels of Figures
\ref{f:phase_ang} and \ref{f:en_ang} shows that the phase-averaged
polarization fraction is significantly more dependent on orientation
than the corresponding phase curves.  This is largely due to the phase dependence of
the polarization angle, and is most clearly illustrated by the solid
black ($\theta_{\rm rot}=45^\circ$, $\theta_{\rm los}=70^\circ$) and
dotted red curves ($\theta_{\rm rot}=70^\circ$, $\theta_{\rm
  los}=45^\circ$) in the bottom panels of Figure~\ref{f:phase_ang}. 
In general, for $\theta_{\rm los} < \theta_{\rm rot}$,
the polarization
angle tends to sweep out the full $180^\circ$ range as the magnetic
axis rotates on the plane of the sky.  A larger polarization angle
variation over a rotational period typically leads to a weaker 
phase averaged
polarization fraction because the Stokes 
Q and U parameters
have different signs at different rotational phases and 
tend to cancel. Indeed, the orthogonal rotator has the highest polarization
fraction because the polarization angle shows the weakest phase
dependence.

The variation of the phase-averaged polarization angle with energy is
also orientation dependent.  In practice, it is difficult to
extract useful constraints from this quantity alone, since it is the
phase dependence of the polarization vector, both angle and magnitude,
that determines the phase-average.  Rather different orientations
can provide very similar rotation of the polarization
angle with energy, while rather similar geometries can give
significantly different results.

\begin{figure}
	\includegraphics*[width=\columnwidth]{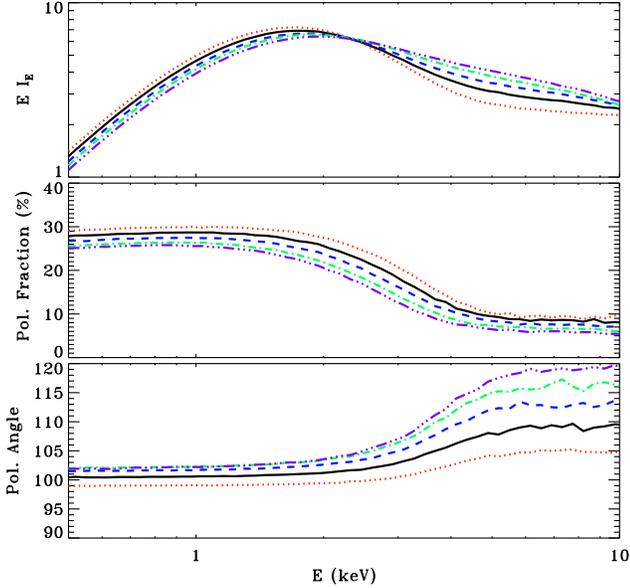}
\caption{
Phase-averaged intensity (top), polarization fraction (middle), and
polarization angle (bottom) as a function of photon energy for models
with different $\Delta \phi_{\rm NS}$.  Magnetic and viewing
orientations corresponding to $\theta_{\rm rot}=45^\circ$ and
$\theta_{\rm los}=70^\circ$ are assumed.  Curves have the same
meaning as in Figure \ref{f:phase_twist}.}
\label{f:en_twist}
\end{figure}

Figure~\ref{f:en_twist} shows the variation of the phase-averaged
observables with twist angle for the same orientation as in Figure
\ref{f:phase_twist}.  Note that there is some diversity in the shape
of the spectra for the total intensity, even though we vary
$\gamma_{\rm max}$ simultaneously with $\Delta \phi_{\rm NS}$ to keep
the high energy spectral slope approximately constant.  As in the
phase-resolved case, the polarization fraction decreases with
increasing $\Delta \phi_{\rm NS}$, but Figure \ref{f:en_twist}
demonstrates that a drop in polarization fraction occurs for all
photon energies.
For increasing energy,
the polarization angle rotates
by an amount which increases monotonically with $\Delta \phi_{\rm
  NS}$.  In principle, this dependence could be used to probe the
twist angle, but the rotation of the polarization angle depends on the
orientation as well, so this needs to be independently constrained.

\begin{figure}
\includegraphics*[width=\columnwidth]{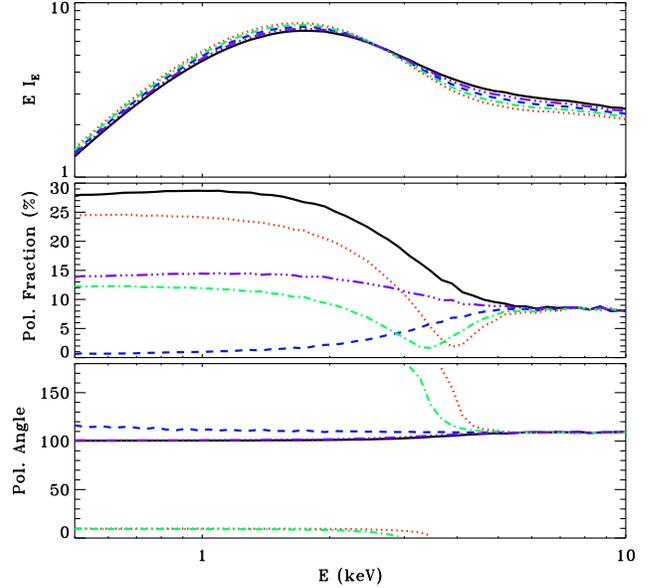}
\caption{
Phase-averaged intensity (top), polarization fraction (middle), and
polarization angle (bottom) as a function of photon energy for models
with different seed photon polarization.  Magnetic and viewing
orientations corresponding to $\theta_{\rm rot}=45^\circ$ and
$\theta_{\rm los}=70^\circ$ are assumed.  Curves have the same
meaning as in Figure \ref{f:phase_seed}.}
\label{f:en_seed}
\end{figure}

Figure~\ref{f:en_seed} compares the phase-averaged observables for
different seed photon distributions, assuming the same orientation as
in Figure~\ref{f:phase_seed}.
The curves correspond 
again 
to linear combinations of models t10\,g20\,c90\,Eu and
t10\,g20\,c90\,Ou.  At low energies ($\lesssim 2$~keV), the unpolarized and
partially polarized seed photon distributions have low polarizations
fractions, as expected.
The pure O-mode distribution has a lower linear polarization than the pure 
E-mode, because the 
small number of scattered photons at low energy with a net E-mode polarization 
which cancels some of
the O-mode seeds.  At higher energies (above $\sim 5$~keV),
which are scattering dominated, mode exchange leads to a common value
for the polarization fraction, independent of the seed distribution.

The variation of the polarization in the transition region from
2-5~keV is very sensitive to the seed photon distribution.  As the
t10\,g20\,c90\,Ou model transitions from being dominated by O-mode seed
photons to E-mode scattered photons, there is a clear drop and
subsequent rise in the polarization, which coincides with a $90^\circ$
rotation of the polarization.  A similar trend is seen in the
partially polarized, but O-mode dominated (75\% O-mode, 25\% E-mode)
model.  In contrast, the E-mode dominated models show a continuous
decline, with only modest rotation of the polarization angle.  Hence,
observations of the polarization fraction as a function of photon
energy in this transition region can
discriminate clearly between
predominately O-mode and predominantly E-mode seed photons
distributions.  
This would help elucidate the mechanism providing the dominant contribution
to the thermal luminosity. 

\begin{figure}
\includegraphics*[width=\columnwidth]{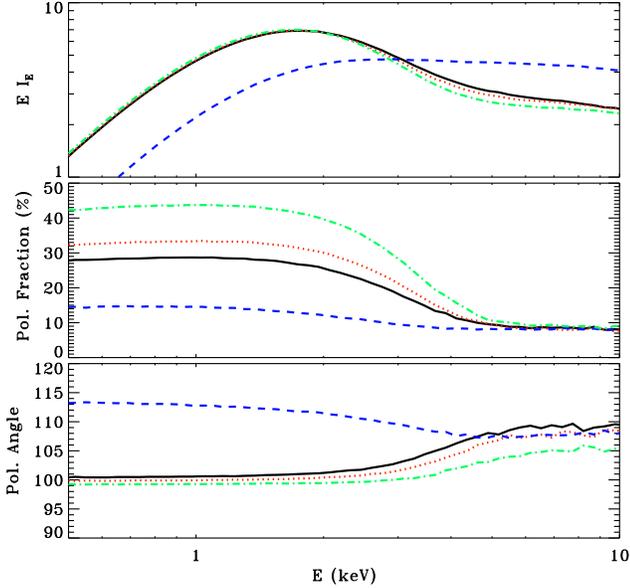}
\caption{
Phase-averaged intensity (top), polarization fraction (middle), and
polarization angle (bottom) as a function of photon energy for various
assumptions about NS surface emission regions and photon propagation.
Magnetic and viewing orientations corresponding to $\theta_{\rm
  rot}=45^\circ$ and $\theta_{\rm los}=70^\circ$ are assumed.  Curves
have the same meaning as in Figure \ref{f:phase_cap}.}
\label{f:en_cap}
\end{figure}

Figure \ref{f:en_cap} shows the effects of confining emission to polar
caps.  At high energies, the models all approach a common value for the
polarization fraction.  The differences in the fraction of scattered
photons primarily drive the observed differences in the polarization
fraction at low energies.  The highly polarized seed photons tend to
escape more easily along the poles, particularly when the effects of
light bending are neglected.  The exception is the model with emission
from only the south pole, because the observer primarily sees
scattered photons for the assumed orientation ($\theta_{\rm
  rot}=70^\circ$, $\theta_{\rm los}=45^\circ$). This also explains the
large difference in the total intensity spectrum.  An observer in the
southern hemisphere (e.g. $\theta_{\rm los}=-45^\circ$) would see
spectra that are qualitatively similar to the other polar cap models.
In contrast to the phase curves of the polarization angle, which were
identical for the different models, the phase-averaged polarization
angles differ due to the variation in the strength of polarization
with rotational phase.

\section{Implications for GEMS}
\label{s:gems}

We now perform simple estimates on the integration times required to obtain polarization
information from magnetar candidates, based on current GEMS sensitivity parameters.
The product of the GEMS effective area and modulation factor
rises from zero at an energy of $2$~keV to its maximum around $3$~keV, and then decreases steadily,
in such a way that by $6$~keV it is smaller by a factor of three from the peak 
[T. Kallman, private communication; see also \citet{swank2010}]. 
This and the typical form of magnetar candidate spectra implies that the bulk of the measured polarization signal 
will come from the transition between thermal and non-thermal parts of the spectrum. In \S\ref{s:results}  we showed
that the polarization fraction in this region can have large variations
depending on the polarization level of the surface emission (and the temperature of the blackbody, which
we do not change in this study).
According to our results, the non-thermal tail is expected to produce a more uniform and
predictable polarization level, albeit at a significantly small flux level. 
Thus we base our
estimates on trying to obtain a statistically significant measurement in this part of the
spectrum, as it is the limiting factor.

Following \citet{weisskopf2010}, we estimate uncertainties by assuming that the polarization
noise follows a Poisson distribution. The $1$-sigma uncertainty in the modulation amplitude
measured by the detector when rotated around the line-of-sight is given by
\begin{equation}
\label{eq:error_amplitude}
\Delta a_{1\sigma} = 2\sqrt{\frac{1}{N}\ln{\left(\frac{1}{1-C}\right)}}\,\bigg|_{C=0.68} = 2.1\times 10^{-2} N_4^{-1/2},
\end{equation}
where $N$ is the number of counts, $N_4 = N/10^4$, and $C$ is the confidence level. The modulation
amplitude produced by the source is
\begin{equation}
\label{eq:modulation_signal}
a_0 = m\,\Pi_L,
\end{equation} 
where $m$ is the modulation of the detector, and $\Pi_L$ is the linear polarization 
fraction (eq.~[\ref{eq:pol_fraction}]). We assume here that the contribution from the background
is not important. Combining equations~(\ref{eq:error_amplitude}) and (\ref{eq:modulation_signal})
allows determination of the number of counts needed to measure $a_0$ to within a given number of sigmas $n_\sigma$,
\begin{equation}
\label{eq:required_counts}
N_{\Pi} = 4\times10^3\left(\frac{n_\sigma}{3} \right)^2\left(\frac{0.5}{m}\right)^2\left(\frac{20\%}{\Pi_L}\right)^2.
\end{equation}
The $68\%$ confidence contour of measured polarization angle corresponding to $\Delta a_{1\sigma}$ is \citep{weisskopf2010}
\begin{equation}
\label{eq:polangle_extrema}
\tan \chi_{1\sigma} =  \frac{a_0\sin\chi_0 + \Delta a_{1\sigma}\sin\psi}{a_0\cos\chi_0 + \Delta a_{1\sigma}\cos\psi},
\end{equation}
where $\chi$ and $\chi_0$ are the measured and true polarization angle (eq.~[\ref{eq:pol_angle_def}]), respectively, and $\psi$ is
a free parameter. Finding the extrema of equation~(\ref{eq:polangle_extrema}) yields 
$\psi_{ext} = \chi_0 \pm \arccos{(-\Delta a_{1\sigma}/a_0)}$. Setting $\chi_0 = 0$, to which the extrema of
equation~(\ref{eq:polangle_extrema}) are weakly sensitive, and keeping terms linear in $\Delta a_{1\sigma}/a_0$ yields
\begin{equation}
\Delta \chi_{1\sigma} = \chi - \chi_0 = \arctan{\left(\frac{\Delta a_{1\sigma}}{a_0}\right)}\simeq \frac{\Delta a_{1\sigma}}{a_0}.
\end{equation}
One can then straightforwardly obtain the number of counts required for a given uncertainty in 
the polarization angle measurement
\begin{equation}
N_\chi = 1.5\times 10^4 \left(\frac{10\textrm{ deg}}{\Delta \chi_{1\sigma}} \right)^2\left(\frac{0.5}{m}\right)^2\left(\frac{20\%}{\Pi_L}\right)^2. 
\end{equation}
In this simple analysis, the ratio $N_\chi/N_\Pi \propto (n_\sigma/\Delta \chi)^2$ is insensitive to the modulation of
the detector
and the polarization fraction of the source. 

\begin{deluxetable*}{lcccccl} 
\tablecaption{Estimated GEMS Count Rates and Observational Times for Selected Magnetar Candidates\label{t:rates}}
\tablewidth{0pt}
\tablehead{
\colhead{Object} &
\colhead{$\dot{N}_{2-10}$ (s$^{-1}$)} &
\colhead{$\dot{N}_{2-4}$  (s$^{-1}$)}&
\colhead{$\dot{N}_{4-10}$ (s$^{-1}$)} &
\colhead{$T^\Pi_{4-10}$  (ks)} &
\colhead{$T^\chi_{4-10}$ (ks)} &
\colhead{Ref.\tablenotemark{a}}
}
\startdata                                        
4U 0142+61            & 3.27  & 3.12  & 0.15  &  27 &  101 & 1\\
1RXS J170849.0-400910 & 0.767 & 0.684 & 0.083 &  48 &  180 & 2\\
1E 1841-045           & 0.557 & 0.475 & 0.082 &  49 &  184 & 3\\
1E 2259+586           & 0.723 & 0.700 & 0.023 & 174 &  653 & 4\\
SGR 1900+14           & 0.113 & 0.095 & 0.018 & 222 &  833 & 5\\ 
\enddata
\tablenotetext{a}{References: (1) \citet{white1996}, (2) \citet{rea2007c}, (3) \citet{morii2003}, (4) \citet{woods2004}, 
(5) \citet{mereghetti2006}}
\end{deluxetable*} 

For a given magnetar candidate, the count rate in a given energy band is
\begin{equation}
\dot N_{E_1 - E_2} = \int_{E_1}^{E_2} \left( \frac{\totd N}{\totd E}\right)\, A_{\rm eff}(E)\, \totd E,
\end{equation}
where $[E_1,E_2]$ are the limits of the energy band, $\totd N / \totd
E$ the photon number flux, and $A_{\rm eff}$ is the effective area of
GEMS. We obtain $\totd N/\totd E$ from published fits to a handful of
bright magnetar candidates in the sample of \citet{kaspi2010}, which
are selected to be in the quiescent stage\footnote{We reconstruct the
  number flux $dN/dE$ using the published fit parameters in XSPEC for the
  best-fit blackbody plus power law models with neutral absorption
  (PHABS) included.  In cases where model normalizations were not reported, we chose
  values to match the reported (energy band integrated) fluxes or flux ratios.
  Since these models provide a reasonable fit for all
  sources under consideration here, they are sufficient for our
  approximate estimates.}. It is well known that some of these sources
experience long term changes in their flux and spectral properties
(e.g., \citealt{kaspi2007}), hence these estimates serve as a rough
reference rather than a detailed prediction.

The resulting count rates in photons per second are shown in Table~\ref{t:rates}. Alongside the
number of counts over the entire sensitivity range of GEMS ($2-10$~keV), we show the rates over
the soft and hard bands employed in \S\ref{s:phase_resolved} ($2-4$~keV and $4-10$~keV, respectively), which contain 
the transition between thermal and non-thermal components, and the pure power-law regime 
in most sources, respectively.
We also show fiducial integration times for the $4-10$~keV frequency band,
\begin{eqnarray}
T^\Pi_{4-10}  & = & \frac{N_\Pi}{\dot{N}_{4-10}}\\
T^\chi_{4-10} & = & \frac{N_\chi}{\dot{N}_{4-10}},
\end{eqnarray}
with $m = 0.5$, $\Pi = 20\%$, $n_\sigma = 3$ and $\Delta \chi_{1\sigma} = 10$~deg. Scaling the numbers with 
respect to this choice of parameters is straightforward. 

Obtaining just one phase-averaged data point in the pure power-law region of the spectrum
requires integration times of the order of tens of kiloseconds or more. Because the ratio of 
count rates in the $2-4$~keV to $4-10$~keV bands is $5-30$, one would
be guaranteed a higher energy resolution below $4$~keV, unless the seed
photons are very unpolarized. Even then, one would be able to detect a
single data point with a $4.4\%$ polarization fraction at 3-$\sigma$ in the 
band $2-4$~keV for 4U 0142+61. An unpolarized source of seed photons would require
a cancellation of the contribution from returning currents and deep cooling (\S\ref{s:seed_distribution}),
which we consider to be a less likely scenario. Thus prospects for obtaining independent clues
about the mechanism producing the non-thermal X-rays are certainly feasible for several of
the brightest AXPs and SGRs.

The observational requirements for phase-resolved polarimetry, on the other
hand, increase linearly with the number of data points per rotation
period, if the same accuracy wants to be maintained. Measurement of the
polarization fraction with 5 points in phase is not very demanding, 
at least for the brightest objects in Table~\ref{t:rates}. The $1\sigma$ uncertainty
in the polarization angle using $T^\Pi_{4-10}$ per phase point is $\Delta \chi_{1\sigma} \simeq 20$~degrees,
which is satisfactory if the spatial scale of the twist is large and the orientation is
favorable (e.g., Figures~\ref{f:phase_ang}, \ref{f:phase_twist}, and \ref{f:phase_seed}). 
Accurately capturing smaller variations in polarization angle
in both spatial scale and amplitude (e.g., lowest panel in Figure~\ref{f:phase_distdir}), 
would become much more demanding,
with count rates easily exceeding $10^6$~s even for 4U 0142+61.

A measure of the polarization angle would be especially useful around outbursts,
as it has already been seen that sources experience persistent changes
in torque and pulse profiles (e.g., \citealt{woods2004,woods2007}). Simultaneous
monitoring with RXTE and high resolution spectra with Chandra or XMM-Newton would be the ideal complement
to deep polarization observations.

\section{Summary and Discussion}

We have studied the effects of resonant cyclotron scattering in a twisted magnetosphere
on the outgoing polarization signal of magnetars. We have employed simple prescriptions for
the magnetic field geometry, particle energy distribution, and seed spectrum. When the pair 
multiplicity is not too large, the dielectric properties of the magnetosphere are dominated
by vacuum polarization out to large distances (eq.~[\ref{eq:dielectric_ratio}]), with the following implications:
\newline

\noindent 1. -- There is an effective separation of the regions where resonant cyclotron
                scattering and polarization eigenmode coupling take place, with the
                consequence that scattering occurs mostly in the adiabatic limit (Fig.~\ref{f:surfaces}).
                Because the light cylinder is safely outside of the zone where polarization
                modes freeze, circular polarization is always less than $\sim 10\%$ (e.g., Figure~\ref{f:trajectory_error}).
                \newline

\noindent 2. -- Regardless of the polarization distribution of surface photons, resonant
                cyclotron scattering by transrelativistic charges introduces a characteristic
                polarization fraction of the order of $\sim 10-30\%$. This number arises
                from the mode exchange probability, dependence of optical depth on polarization,
                multiple scattering, and geometry of the system (\S\ref{s:corotating}). The transition
                from surface-dominated polarization to scattering-dominated takes place over a
                region of the spectrum extending over the interval $\sim 2-5$~keV for a thermal
		spectrum with $kT = 0.4$~keV (Figures~\ref{f:en_ang}-\ref{f:en_cap}).
                Measurement of phase-averaged polarization fraction
                allows an unambiguous identification of resonant comptonization as the mechanism
                producing the non-thermal soft X-ray emission. This is particularly relevant for
		sources such as XTE J1810-197, whose spectrum can be fit with a comptonized
		model or two-temperature blackbody \citep{halpern2005}. 
                \newline

\noindent 3.    Phase averaged measurements can also discern
		the dominant polarization mode of seed photons (Fig.~\ref{f:en_seed}), allowing an
		independent constraint on the mechanism providing the thermal luminosity. A dominance
                of E-mode photons in the thermal spectrum would favor deep cooling of the neutron
                star \citep{thompson1996}, setting constraints on magnetized atmosphere models 
                (e.g., \citealt{vanadelsberg2006}).
                An O-mode dominance in the unscattered component would indicate thermal
		emission from returning currents hitting the stellar surface \citep{thompson2002}.
                \newline

\noindent 4. -- Measurement of the phase-resolved polarization angle
                allows measurement of the spatial extent of the twist angle
                at radii $\sim 100R_{\rm NS}$. 
                Because the mode exchange probability always favors the E-mode regardless of 
                system parameters (eq.~[\ref{eq:mode_switch_prob}]), scattered photons are 
                predominantly in this linear eigenstate 
                and their adiabatic propagation keeps track of the magnetic field geometry, 
                with little sensitivity
                to input parameters. Photons remain mostly linearly polarized as eigenmodes
   		couple, a direct consequence of the dominance of vacuum polarization and 
                the slow rotation of magnetars. For each photon, the polarization vector preserves its orientation
                relative to the direction of the magnetic field at the point of freezing.
                In the case of a globally twisted dipole,
		it may even be possible to obtain the geometric parameters of the system
		by inverting the polarization angle as a function of phase. We have not 
		explored this possibility (e.g., lower panels of Figures~\ref{f:phase_ang} 
                and~\ref{f:polangle_notwist}).
		It is worth keeping in mind that this effect is dependent on magnetospheric
                scattering only in that
		a particular eigenmode is preferred: in principle it is also present in the
                unscattered component (e.g., \citealt{heyl2000}). 
                \newline

\noindent 5. -- The asymmetries in the angular distribution of scattered radiation are due 
                to the combination of charge motion along field lines and relativistic 
                aberration. This also imparts an asymmetry in the polarization fraction.
                If scattering charges consist predominantly of electrons moving from 
                one magnetic pole to the other, then a north-south asymmetry is introduced
                in the polarization fraction. Phase-resolved polarization fraction 
                will display a peak displaced by one-half of a phase relative to the
                peak in intensity if the particle energy distribution is uni-directional (Figure~\ref{f:phase_distdir}).
                If the magnetosphere is dominated by pairs at these radii ($\sim 10R_{\rm NS}$),
                then two symmetric peaks are expected in both the polarization distribution
                and the radiation flux, barring a strong asymmetry in the surface emission.
                \newline

\noindent 6. -- Based on the expected sensitivity of GEMS and the spectra of a few bright
                magnetars, we estimate that phase-averaged polarimetry requires several
                tens to a few hundred kilosecond exposures per source for obtaining
                at least one data point in the non-thermal tail of the spectrum (Table~\ref{t:rates}). 
                Phase-resolved polarimetry increases the requirements proportional to the number
                of points in phase for fixed accuracy. Detailed measurements of changes
                in the polarization angle as a function of phase seem feasible only for the brightest sources,
                and would require integration times approaching or exceeding a $10^6$~s.
                \newline

Our analysis ignores the contribution of the magnetospheric plasma to the dielectric tensor.
This is a good approximation if the density of charge carriers satisfies $n_e\sim J/(ec)$, 
as found in \citet{beloborodov2007b}. That study, however, focused on the inner part of
the closed magnetosphere at radii $r\sim R_{\rm NS}$, where a number of effects present
at larger radii are absent. 
Pair cascades in the closed field circuit could be triggered by energy deposition from current driven
instabilities close to the star \citep{thompson2008a}. A significant number of pairs at $\sim 100R_{NS}$
then alters the relative contribution of plasma and vacuum to the dielectric tensor (eq.~[\ref{eq:dielectric_ratio}]).
When the plasma component is non-negligible, and $\omega \gg \omega_c$, propagation eigenmodes become
increasingly circular as the charge density is increased, for most propagation directions (e.g., \citealt{meszaros1992}).
Capturing this effect, particularly if the transition from vacuum to plasma dominated modes occurs
around $r_{\rm pl}$, requires a more careful consideration of the distribution of currents in the
magnetosphere than our simple space-independent particle energy distribution. We have opted for
a self-consistent (but potentially incomplete) first step in modeling this problem.

We have also ignored the contribution of scattering by ions close to the star (e.g., \citealt{thompson2002}).
Since the dielectric properties of the plasma would not change relative to the case analyzed here, we expect
only quantitative changes relative to scattering by uni-directional electrons. Ion scattering would still be in the adiabatic
limit, and modes would freeze at the same distance from the star as in the case of photons scattered by electrons. 
Hence, we would not expect significant changes in the phase dependence of the polarization angle.
Changes would mostly affect the intensity and polarization fraction, and would be 
due to general relativistic effects (in photon propagation and magnetic field, the second of which we are ignoring) 
as well as due to reflection or absorption
by the star, whose radius would now be comparable to the resonant surface. And obviously photons that escape the ion resonance surface
can still be subject to scattering by electrons at larger radii, so the net observed anisotropies could increase. 

Most of our calculations assume the simplest possible prescription for the seed photons: a blackbody of uniform
surface temperature, $100\%$ polarized, and isotropically emitting, in order to isolate the
magnetospheric imprint on the polarization signal. Out of these assumptions, we
expect those concerning the surface temperature and beaming distribution to potentially alter
our results once they are relaxed.
To gauge the effects of a non-uniform surface temperature, we have explored the extreme case 
of narrow polar caps around one or both magnetic poles, motivated by results from thermal conduction
calculations with a strong internal toroidal field \citep{perez-azorin2006}. We find 
minor differences relative to the
uniform surface temperature distribution unless one polar cap emits substantially more 
than the other (Figs.~[\ref{f:phase_cap}] and [\ref{f:en_cap}]).
A radiation intensity that has a non-trivial distribution around the local magnetic field 
(e.g., \citealt{vanadelsberg2006}) could also distort
the signature due to the directionality of particles. We have not addressed this issue here,
as it would require a realistic atmosphere for self-consistency. 

Given our assumptions, the polarization angle is an almost perfect tracer of the magnetic field
geometry at $\sim 100R_{\rm NS}$ from the star. In this study we have employed a very
simple field geometry, which is self-similar and thus having a local direction
independent of radial distance from the star. In a more realistic magnetosphere,
the twist angle is not necessarily global, and evolves both in time and space 
\citep{thompson2002,beloborodov2007b,thompson2008a,beloborodov2009}. 
Measurements of the magnetospheric twist at large radii would be most useful
during and following periods of burst activity, where the magnetosphere is expected
to undergo a significant rearrangement, as shown by changes in the pulse profiles
before and after outbursts (e.g., \citealt{woods2004,woods2007}). 
A pulse profile with
strong asymmetries and small-scale structure, if associated with localized twists,
should be concurrent with a phase-resolved polarization angle that displays high temporal 
frequency and small amplitude components.

\acknowledgements
We thank Chris Thompson for stimulating discussions and comments on the manuscript.
We also thank Tim Kallman and the GEMS team for providing instrument sensitivity
data. Comments by an anonymous referee improved the presentation of the paper. 
R.~F. is supported by NASA through Einstein Postdoctoral Fellowship
grant number PF-00062, awarded by the Chandra X-ray Center, which is operated
by the Smithsonian Astrophysical Observatory for NASA under contract NAS8-03060.
S.~W.~D. is supported by NSF AST-0807432, NASA NNX08AH24G, NSF AST-0807444, and
in part by NSERC of Canada.
Computations were performed at the IAS Aurora and CITA Sunnyvale clusters.
The latter was funded by the Canada Foundation for Innovation.

\appendix

\section{Coordinate Transformations}
\label{s:rotations}

\begin{figure*}
\includegraphics*[width=\textwidth]{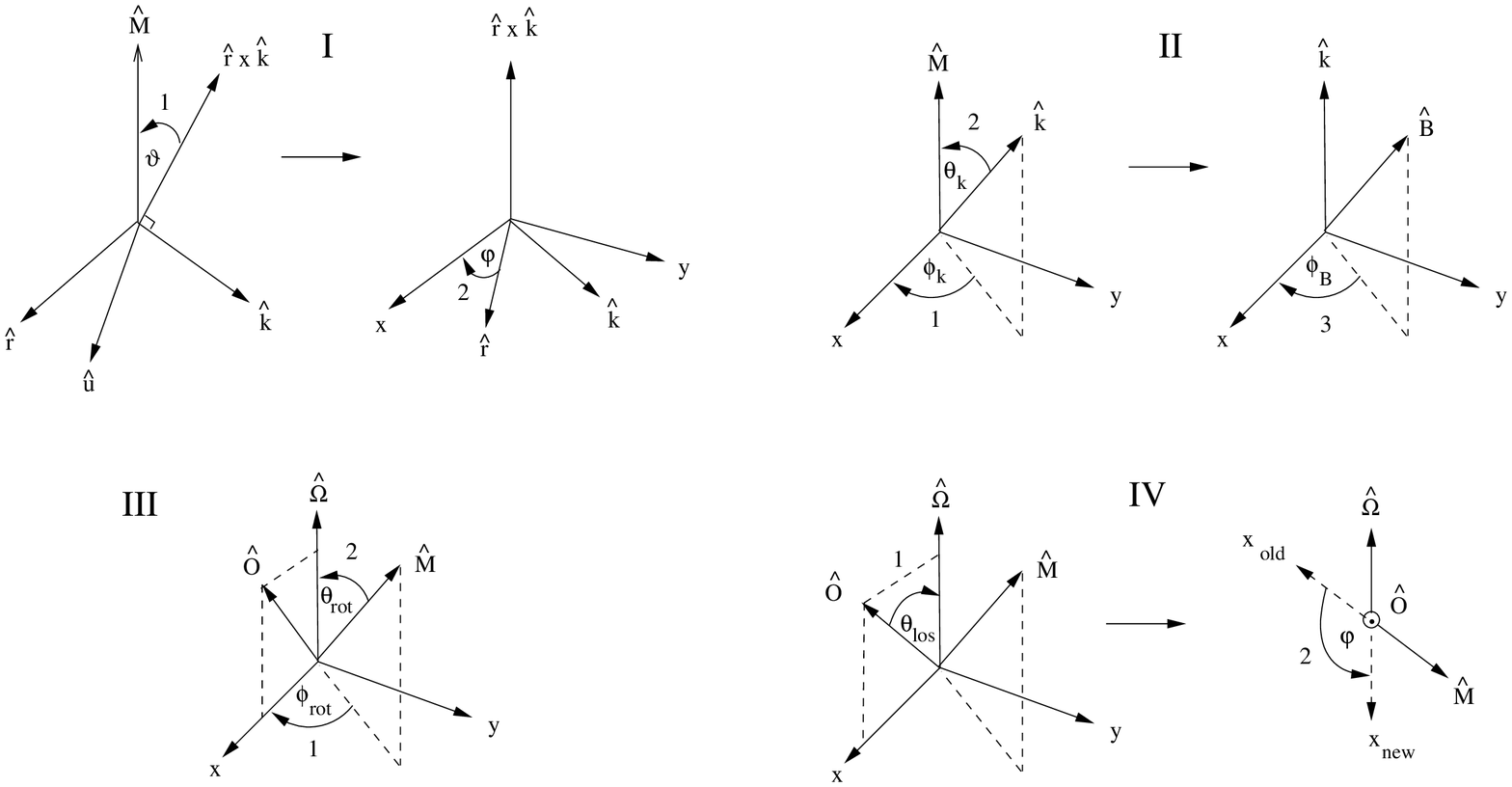}
\caption{Coordinate systems and geometric transformations employed in the Monte Carlo calculation, 
labeled by roman numbers. The vectors $\hat M$, $\hat \Omega$, and $\hat O$ denote the magnetic axis, rotation axis, 
and line-of-sight (see also eqns.~[\ref{eq:murot}] and [\ref{eq:mulos}]). The arabic numerals indicate the order in which
each rotation matrix is applied (see text).}
\label{f:coordsys}
\end{figure*}

We present here all the coordinate transformations employed in the calculation. We denote the
coordinate system with the magnetic axis $\hat M$ in the $z$-direction by the subscript $M$.
Each alternative coordinate system employed is labeled by roman number subscripts on
vectors.
Each of them is illustrated in Figure~\ref{f:coordsys} and explained below.
For conciseness, we write the transformations in terms of the elementary rotation matrices
\begin{equation}
\label{eq:rot_matrices}
R_x(\vartheta) = \left[\begin{array}{ccc} 1 & 0 & 0\\ 0 & \cos\vartheta & -\sin\vartheta\\ 
                                       0 & \sin\vartheta & \cos\vartheta\end{array}\right]\qquad
R_y(\vartheta) = \left[\begin{array}{ccc} \cos\vartheta & 0 & \sin\vartheta\\ 0 & 1 & 0\\ 
                                       -\sin\vartheta & 0 & \cos\theta\end{array}\right]\qquad
R_z(\vartheta) = \left[\begin{array}{ccc} \cos\vartheta & -\sin\vartheta & 0\\ \sin\vartheta & \cos\vartheta & 0\\ 
                                                             0 & 0 & 1\end{array}\right],
\end{equation}
which when applied to a column vector $\mathbf x$ produce a counterclockwise rotation by an angle $\vartheta$ around each of 
the cartesian coordinate axes, respectively. We also make use of the more general angle-axis formula: given a unit vector $\hat u$,
the rotation matrix that produces a counterclockwise rotation by an angle $\vartheta$ around $\hat u = u_x\hat x + u_y\hat y
+ u_z\hat z$ is
\begin{equation}
R_u(\vartheta) = \left[\begin{array}{ccc} 
u_x^2 + (1-u_x^2)\cos\vartheta & u_x u_y (1-\cos\vartheta) -u_z\sin\vartheta & u_x u_z (1-\cos\vartheta) + u_y\sin\vartheta\\
u_x u_y (1-\cos\vartheta) + u_z\sin\vartheta & u_y^2 + (1-u_y^2)\cos\vartheta &  u_y u_z (1-\cos\vartheta) - u_x\sin\vartheta\\
u_x u_z (1-\cos\vartheta) - u_y\sin\vartheta & u_y u_z (1-\cos\vartheta) + u_x\sin\vartheta & u_z^2 + (1-u_z^2)\cos\vartheta
\end{array} \right],
\end{equation}
with (\ref{eq:rot_matrices}) corresponding to $\hat u = \hat x$, $\hat y$, and $\hat z$, respectively.

\subsection{Light Bending (I)}

Since a photon orbit in a Schwarzschild metric has two conserved quantities (energy and angular momentum), one can
map it to the $\theta = \pi/2$ plane in spherical polar coordinates without loss of generality (e.g., \citealt{schutz2009}). 
We choose the $x$-axis to be coincident with the position vector $\hat r$, and the $\hat z$ axis along $\hat r \times \hat k$.
The photon wave vector is transformed according to
\begin{equation}
\label{eq:transform_bend_1}
\hat k_I = R_z(-\varphi)\,R_u(\vartheta)\, \hat k_M.
\end{equation}
The first rotation around $\hat u = (\hat r \times \hat k) \times \hat M / |\hat r \times \hat k|$, with
$\cos\vartheta = (\hat r \times \hat k)\cdot \hat M / |\hat r \times \hat k|$, puts the trajectory
on to the equatorial plane. 
We then rotate the radial vector $\hat r$ towards $\hat x$,
with 
\begin{equation}
\tan\varphi = \frac{[R_u(\vartheta)\hat r_M]_y}{[R_u(\vartheta)\hat r_M]_x}. 
\end{equation}
The normalized wave vector evolved in 
equations~(\ref{eq:light_bend_r})-(\ref{eq:light_bend_s}) 
corresponds to the components of $\hat k_I$ in spherical polar coordinates.
The transformation is ill-defined for radial propagation, which is nonetheless unaffected by
light bending. In practice, we revert to planar propagation whenever $1- |\hat k \cdot \hat r| < 10^{-6}$.

\subsection{Polarization Evolution (II)}

We start each integration of equations~(\ref{eq:prop_ax_general})-(\ref{eq:prop_ay_general}) 
in a frame such that the wave vector $\hat k$ points in the $z$ direction and the magnetic
field lies in the $x-z$ plane. The transformation for the electric field vector $\mathbf A$ is
\begin{equation}
\mathbf A_{II} = R_z(-\phi_B)\,R_y(-\theta_k)\, R_z(-\phi_k)\, \mathbf A_M,
\end{equation}
where $\theta_k$ and $\phi_k$ are the polar and azimuthal angle of the wave vector relative
to the magnetic axis, and
\begin{equation}
\tan\phi_B = \frac{[R_y(-\theta_k)\, R_z(-\phi_k)\,\hat B]_y}{[R_y(-\theta_k)\, R_z(-\phi_k)\,\hat B]_x}.
\end{equation}

\subsection{Rotational Phase (III)}

Our coordinate system for phase-dependent quantities is such that the rotational axis $\hat \Omega$ points in the $z$-direction, and
at phase zero both the magnetic axis $\hat M$ and the direction to the line-of-sight $\hat O$ are in the $x-z$ plane, 
at angles $\theta_{\rm rot}$
and $\theta_{\rm los}$ from $\hat \Omega$, respectively. 
What is needed are the coordinates of the line-of-sight $\hat O$ relative
to the magnetic axis, so that stored quantities are retrieved to construct phase curves. The transformation is
\begin{equation}
\hat O_M = R_y(-\theta_{\rm rot})\, R_z(-\phi_{\rm rot})\, \hat O_{III}, 
\label{eq:rotphase}
\end{equation}
with $\hat O_{III} = \sin\theta_{\rm los}\hat x_{III} + \cos\theta_{\rm los}\hat z_{III}$, and $\phi_{\rm rot}$ the
rotational phase.

\subsection{Stokes Parameters (IV)}

A rotation of the polarization plane around the wave vector affects only Q and U \citep{chandrasekhar1960}.
Counterclockwise rotation by an angle $\varphi$ yields
\begin{eqnarray}
Q^\prime & = & Q \cos 2\varphi + U\sin 2\varphi\label{eq:stokesrot}\\
U^\prime & = & -Q\sin 2\varphi + U\cos 2\varphi.
\end{eqnarray}
For storage after photon escape to infinity, we take the $x$ axis in the $\hat k - \hat M$ plane. For
phase dependent quantities, however we rotate this axis to coincide with the $\hat k - \hat \Omega$ plane.
The needed angle $\varphi$ is obtained by rotating $\hat O_{III}$ so that it lies along the $\hat z$ axis, and
then computing the azimuthal angle between $-\hat M_{IV}$ and $-\hat \Omega_{IV}$, thus
\begin{equation}
\tan (\pi - \varphi) = \frac{[R_y(-\theta_{\rm los})\, \hat M_{III}]_y}{[R_y(-\theta_{\rm los})\,\hat M_{III}]_x},
\label{eq:stokesang}
\end{equation} 
where $\hat M_{III} = R_z(\phi_{\rm rot})\, R_y(\theta_{\rm rot})\, \hat z$ (eq.~[\ref{eq:rotphase}]).

\bibliographystyle{apj}
\bibliography{magnetar,apj-jour}

\end{document}